%% file: main.tex
\documentclass[
    aps,
    pra,
    reprint,
    superscriptaddress
]{revtex4-2}

\usepackage{graphicx}
\usepackage{dcolumn}
\usepackage{bm}
\usepackage{physics}
\usepackage{mathtools}
\usepackage{xcolor}
\usepackage{booktabs}
\usepackage{multirow}
\usepackage{makecell}
\usepackage{enumitem,amssymb}
\usepackage{xcolor}
\newlist{todolist}{itemize}{2}
\setlist[todolist]{label=$\square$}
\usepackage{pifont}

\newcommand{\ignore}[1]{}

\usepackage[hidelinks]{hyperref}
\hypersetup{
    colorlinks=true,
    linkcolor=blue,
    filecolor=blue, 
    urlcolor=blue,
    citecolor=blue
    }

\begin{document}

\preprint{APS/123-QED}

\title{ A Dual Metastable-State Encoding Architecture for Quantum Processing with $^{171}$Yb Atom Arrays}

\author{Chun-Wei Liu} \thanks{These authors contributed equally to this work.} 
\affiliation{Department of Electrical Engineering and Computer Science, University of Michigan, Ann Arbor, Michigan 48109, USA} 

\author{Saiwei Nie} \thanks{These authors contributed equally to this work.} 
\affiliation{Department of Electrical Engineering and Computer Science, University of Michigan, Ann Arbor, Michigan 48109, USA}

\author{Eesha Banerjee} \thanks{These authors contributed equally to this work.} 
\affiliation{Department of Electrical Engineering and Computer Science, University of Michigan, Ann Arbor, Michigan 48109, USA}

\author{Micah Davidson}
\affiliation{Department of Electrical Engineering and Computer Science, University of Michigan, Ann Arbor, Michigan 48109, USA}

\author{Nick Reynolds}
\affiliation{Department of Electrical Engineering and Computer Science, University of Michigan, Ann Arbor, Michigan 48109, USA}

\author{Alyssa L. Miller}
\affiliation{Applied Physics Program, University of Michigan, Ann Arbor, Michigan 48109, USA}

\author{Alex P. Burgers}\email{aburgers@umich.edu}
\affiliation{Department of Electrical Engineering and Computer Science, University of Michigan, Ann Arbor, Michigan 48109, USA}
\affiliation{Quantum Research Institute, University of Michigan, Ann Arbor, Michigan 48109, USA}

\input{abstract}
\maketitle
\input{introduction}
\input{physics_layer}

\input{pipelined_quantum_processor_architecture}
\input{conclusion}
\input{acknowledgement}
\input{appendix}

\bibliography{reference}

\end{document}

%% file: abstract.tex
\begin{abstract}
Neutral-atom arrays combine scalable qubit registers, long coherence times, flexible optical control, and strong, Rydberg-mediated entangling interactions, making them a promising platform for quantum information processing. However, physical error rates remain a challenge, and fault-tolerant quantum error correction (QEC) requires repeated mid-circuit measurement and reset of ancilla qubits without disturbing nearby data qubits. This requirement introduces significant control and architectural overhead, making qubit encoding an important architectural decision. Here, we propose a dual metastable-state qubit encoding for $^{171}\mathrm{Yb}$ atoms that utilizes two independent qubit subspaces in the $(6s6p)\,{^3}\mathrm{P}_0$ and $(6s6p)\,{^3}\mathrm{P}_2$ manifolds. The ${^3}\mathrm{P}_0$ manifold provides a long-coherence nuclear spin (NS) qubit suitable for storage and arithmetic operations, while the ${^3}\mathrm{P}_2$ manifold provides a hyperfine-split (HF) qubit, with $\Delta_{\mathrm{HF}} = 2\pi\times 6.7$ GHz, that enables fast Raman operations and direct state-selective imaging. Coherent shelving between the two metastable manifolds connects the qubit subspaces, allowing operations to be assigned to spectrally distinct processor zones. We simulate single-qubit and two-qubit gate fidelities in ${^3}\mathrm{P}_2$, as well as coherent shelving between the HF and NS qubit subspaces. We incorporate these physical-level estimates into an architectural resource estimation and logical-level simulation. Our approach integrates mid-circuit measurements and fast qubit operations within a single-species platform, providing a versatile framework for future fault-tolerant quantum computing with neutral-atom qubits.

\end{abstract}

%% file: introduction.tex
\section{Introduction}\label{section:introduction}
\begin{figure*}[!ht]
    \centering
    \includegraphics[width=\textwidth]{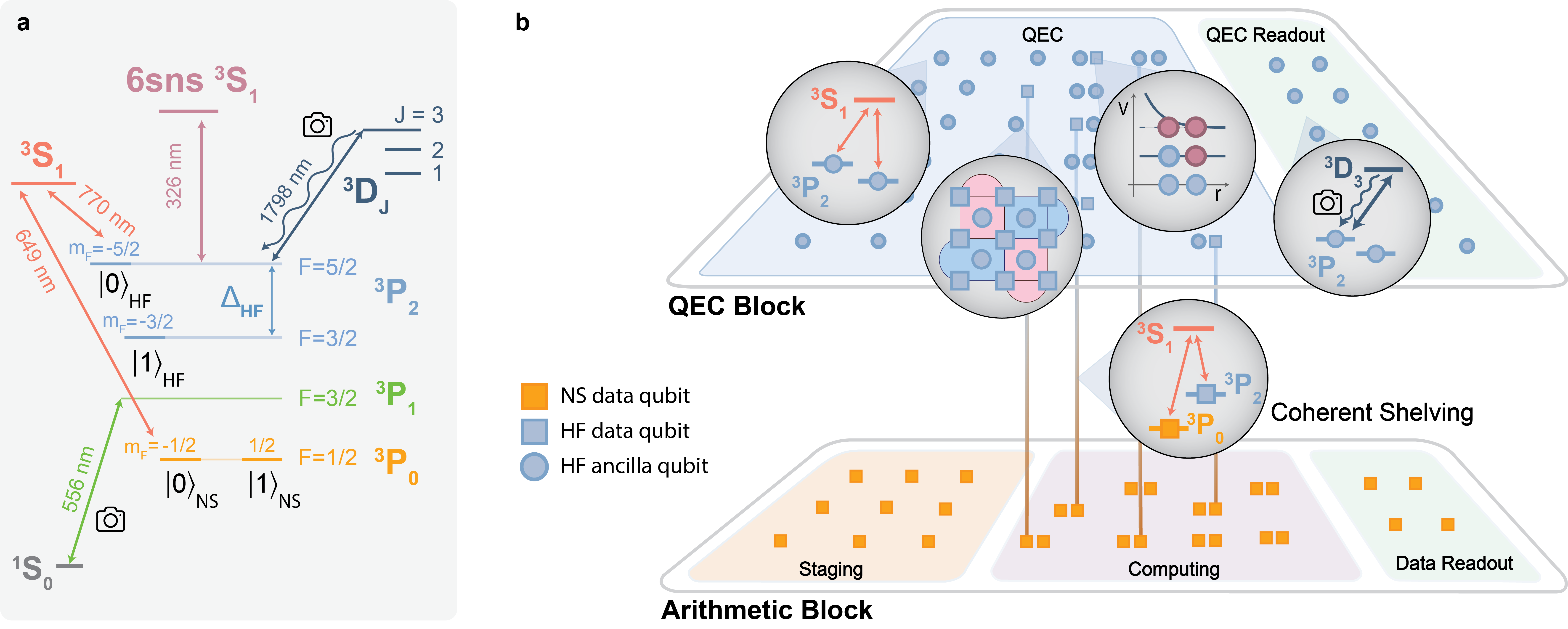}
    \caption{\textbf{Dual-metastable state qubit encoding in a neutral atom processor.} \textbf{a.} Qubits are encoded in the two metastable manifolds of $^{171}$Yb: hyperfine (HF) qubits in the $(6s6p)\,{^3}\mathrm{P}_{2}$ manifold and nuclear spin (NS) qubits in the $(6s6p)\,{^3}\mathrm{P}_{0}$ manifold. Both manifolds have single-photon coupling to the $(6s\mathrm{n}s)\,^3\mathrm{S}_{1}$ Rydberg manifold. The $(6s7s)\, ^3\mathrm{S}_{1}$ manifold provides an intermediate state for single-qubit operations in the HF subspace and coherent shelving to the NS subspace. The $(6s5ds)^3\mathrm{D}_{3}$ manifold provides a non-destructive readout channel via a $1798$ nm cycling transition. \textbf {b.} Processor architecture based on this encoding: the arithmetic block utilizes NS qubits for algorithmic operations, and the QEC block uses HF qubits for QEC protocols. Coherent shelving between the $^3\mathrm{P}_{2}$ and $^3\mathrm{P}_{0}$ manifolds connects the qubit subspaces, allowing data qubits (DQs) to be transferred from the NS to HF subspaces and combined with ancilla qubits (AQs) in the HF subspace for syndrome extraction.}
    \label{fig:fig1}
\end{figure*}

Neutral atoms trapped in arrays of optical tweezers are a versatile and scalable platform for quantum information processing~\cite{Schlosser2001Subpoisson, Bernien2017AtomArray, Manetsch2025Cs6100array, Aaron2026Metasurface, Schymik2021Lifetime6000s, Chiu2025ContineousLoading, li2025fastcontinuouscoherentatom, Manetsch2025Cs6100array}. Qubit basis states are encoded in the internal degrees of freedom of individual atoms, and coherent single-qubit operations are achieved through external control fields. Two-qubit entangling gates are implemented via excitation to Rydberg states, where strong, long-range interactions mediate entanglement. Advancing quantum processors toward practical applications requires scalable platforms and high-fidelity operations, particularly for emerging approaches to fault-tolerant quantum computing~\cite{Rodriguez2025MagicStateDissalation}. Recent experiments in neutral atom arrays have established several capabilities central to scalable, high-fidelity quantum processors. These include high-fidelity single-qubit gates and Rydberg-mediated two-qubit operations~\cite{MaAlex2022YbUniversalGate, Evered2023HighFidelityEntanglingGate, Tsai2025CZLinearResponse}, coherent rearrangement with long-range connectivity that enables quantum error correction protocols~\cite{Kitaev1997QuantumErrorCorrection,Manuel2016Rearranging, Daniel2016Rearranging, Levine2019LevinePichlerGate, Bluvstein2022EntanglingTransport, Manetsch2025Cs6100array}, and photonic link capabilities for modular networking and large-scale entanglement generation~\cite{kuzmich2005atomphotonentanglement, Luan2020PhotonicCrystalTrap, Menon2024PhotonicCrystalTrap, yiyi2024ModularQPU, Li2025InterconnectNode, Sinclair2025OpticalInterconnect, Sunami2025FiberInterconnect}. These features provide the critical components for realizing robust, error-corrected quantum processing of deep circuits using neutral atom arrays~\cite{Bluvstein2024QPU, Bluvstein2025UiversalQPU}.

Atomic platforms support a variety of physical qubit encodings that determine the gate speed, operational control systems, and coherence of quantum gate operations. In alkali atoms, qubits are often encoded in the ground-state hyperfine manifold, where large energy splittings enable fast single-qubit rotations through either microwave or stimulated Raman transitions ~\cite{jones2007fasthyperfinequbit,saffman2010quantuminformation,wang2014quantumCesiumHyperfineQubit}. Alkaline-earth-like atoms (AEAs) expand the design space, offering several qubit encodings within the atoms' ground and metastable manifolds. Nuclear-spin qubits in the ground state and $J=0$ metastable states of fermionic isotopes of strontium ($^{87}$Sr, $I=9/2$) and ytterbium ($^{171}$Yb, $I=1/2$) offer robustness against dephasing due to the absence of hyperfine coupling~\cite{noguchi2011quantum1S0coherenceNuclearSpin,ludlow2015opticallatticeNuclearSpin, Cooper2018SrArray, Saskin2019Ybqubit}. Optical clock qubits exploit the narrow, long-lived transitions between $J=0$ states to provide highly coherent storage with spectrally selective optical control~\cite{Lis2023MidcircuitOMG, Madjarov2020OpticalQubit, Young2020clockqubit, Allcock2021omgIon}. Recent experimental demonstrations with fine-structure qubits in AEA metastable manifolds have established large electronic energy splittings as a useful resource for spectrally selective encoding \cite{Pucher2024FineStructureSr,unnikrishnan2024coherent}. Beyond electronic encodings, neutral atoms can also exploit motional degrees of freedom, providing bosonic modes for quantum information processing~\cite{Adam2025ErasureCoolingMotionalState, li2026quantumsciencearraysmetastable}. This diversity of encodings makes AEAs especially attractive for architectures that assign different processor functions to different internal states.

With current physical error rates in neutral atoms~\cite{evered2026highfidelityentanglinggatesnonlocal, Radnaev2025UnivQPULocalDressing, zhang2025leveragingerasureerrorslogical}, as circuit depth of quantum algorithms increases, accumulated gate errors and decoherence eventually limit computational fidelity ~\cite{evered2026highfidelityentanglinggatesnonlocal, Radnaev2025UnivQPULocalDressing, zhang2025leveragingerasureerrorslogical,Preskill2018quantumcomputingin, Preskill2025MegaGroup}. While optimized compilation and execution can partially reduce circuit overhead~\cite{draper2004LogDepthOptimzation}, mid-circuit measurement is critical for effective error correction in neutral atoms~\cite{Sahay2023ErasureMakeHighTresholdCodes}. Stabilizer-based QEC codes rely on repetitive syndrome extraction to read out error information from data-carrying qubits without directly measuring, and thus disturbing, the encoded quantum state~\cite{Steane1996SteaneCode, gottesman1997stabilizercodesquantumerror, Fowler2012SurfaceCode}. Such non-destructive access can be achieved by introducing an additional set of ancilla qubits (AQs) which are spatially or spectrally isolated from data qubits (DQs) during computation~\cite{Kjaergaard2020SCqubitConcept, Deist2022CavityMidcircuit, Norcia2023Midcircuit}. It is therefore beneficial to separate AQs and DQs into distinct qubit subspaces in dual-type encodings, either across different atomic species~\cite{anand2024dualspiciesBernien, petrosyan2024fastDualSpeciesSaffman, zhang2025dualspeciesAQDQ, white2026quantumcellularautomatadualspecies}, between different isotopes~\cite{zeng2017entanglingDualIsotope, nakamura2024hybridDualisotopeAQDQ}, or within separate manifolds of the same atomic species~\cite{yang2022dualtypeion, Lis2023MidcircuitOMG} to minimize latency associated with atom movement. With a  variety of possible encodings, the metastable states of AEAs are well suited for mid-circuit measurement and error correction ~\cite{Wu2022ErasureConversionAEA,Singh2023Midcircuit,Graham2023Midcircuit, Norcia2023Midcircuit, Ma2023YbMidcircuitErasure, zhang2025leveragingerasureerrorslogical}.

In this paper, we propose a dual-manifold qubit encoding that exploits the unique metastable structure of $^{171}\mathrm{Yb}$ to establish two distinct qubit subspaces within the $(6s6p)\,{^3}\mathrm{P}_0$ and $(6s6p)\,{^3}\mathrm{P}_2$ manifolds. We aim to utilize the particular strengths of the ${^3}\mathrm{P}_2$ manifold, where the large hyperfine splitting ($\Delta_{\mathrm{HF}}=2\pi\times6.7~\mathrm{GHz}$) between the $F=5/2$ and $F=3/2$ states enables fast single-qubit rotations via optical Raman transitions. By using the stretched Zeeman states of the manifold, we can achieve direct, state-selective imaging through the cycling transition $\ket{(6s6p)\,{^3}\mathrm{P}_2,\, F = 5/2,\, m_F = -5/2} \leftrightarrow \ket{(6s5d)\,^3\mathrm{D}_3, F=7/2, m_F=-7/2}$ to perform qubit readout. In the ${^3}\mathrm{P}_0$ manifold, nuclear-spin-1/2 states $\ket{(6s6p)\,{^3}\mathrm{P}_0,\, F = 1/2,\, m_F = \pm 1/2}$ exhibit ultra-long coherence times~\cite{Jenkins2022Ybqubit}, and the coherent control of these states has been well studied in both rf-driven~\cite{Ma2023YbMidcircuitErasure} and optical-driven schemes~\cite{Lis2023MidcircuitOMG}. Combining the strengths of both manifolds, as shown in Fig.~\ref{fig:fig1}a, we present a scheme in which hyperfine (HF) qubits encoded in the ${^3}\mathrm{P}_2$ manifold ($\ket{0}_{\mathrm{HF}}$ and $\ket{1}_{\mathrm{HF}}$) are used for error correction operations, while nuclear-spin (NS) qubits in the ${^3}\mathrm{P}_0$ manifold ($\ket{0}_{\mathrm{NS}}$ and $\ket{1}_{\mathrm{NS}}$) are used for arithmetic operations, enabling mid-circuit functionality and coherence over numerous QEC cycles.

 Under this dual-encoding scheme, the ability to coherently switch between the ${^3}\mathrm{P}_0$ and ${^3}\mathrm{P}_2$ manifolds is critical for our architecture. We simulate coherent transfer using a two-photon Raman transition mediated by the $(6s7s)\,^3\mathrm{S}_1$ state. We note that a similar coherent transfer scheme has recently been demonstrated in fine-structure qubits in bosonic isotopes of AEAs~\cite{tao2025universalgatesmetastablequbit, li2025fastcontinuouscoherentatom}. By leaving the absolute ground state outside the computational subspace, this encoding also supports erasure conversion to detect loss during qubit operations~\cite{Wu2022ErasureConversionAEA, Scholl2023ErasureRydbergSimulator, Adam2025ErasureCoolingMotionalState, zhang2025leveragingerasureerrorslogical}.
 
 Motivated by mid-circuit syndrome measurement, we develop a processor architecture that is partitioned into two operational blocks, an arithmetic block and a QEC block, each encoded in one of the metastable manifolds in our scheme (Fig.~\ref{fig:fig1}b). This design links physical qubit properties directly to architectural function, making co-optimization across the qubit and processor levels essential for scaling neutral-atom systems. We therefore simulate both hardware-level parameters and circuit-level performance metrics to guide and assess the architecture design.

The paper is organized as follows: In Sec.~\ref{section:qubit encoding and trapping}, we introduce the encoding spaces together with their initialization and trapping conditions. In Sec.~\ref{section:universal gate set}, we present simulations of the universal gate operations, including single-qubit operations and coherent transfer obtained from a full-state master-equation solver, and discuss the implementation of state readout. Furthermore, we discuss two-qubit operations in Rydberg states and demonstrate a controlled-phase gate. Finally, in Sec.~\ref{section:pipelined processor}, we combine these encoding attributes into an architecture design that is compatible with mid-circuit syndrome measurement and evaluate its performance on benchmark circuits.

%% file: physics_layer.tex
\section{Dual-metastable State Qubit Encoding and Trapping}\label{section:qubit encoding and trapping}

\begin{figure}[!ht]
\centering
    \includegraphics[width=\columnwidth]{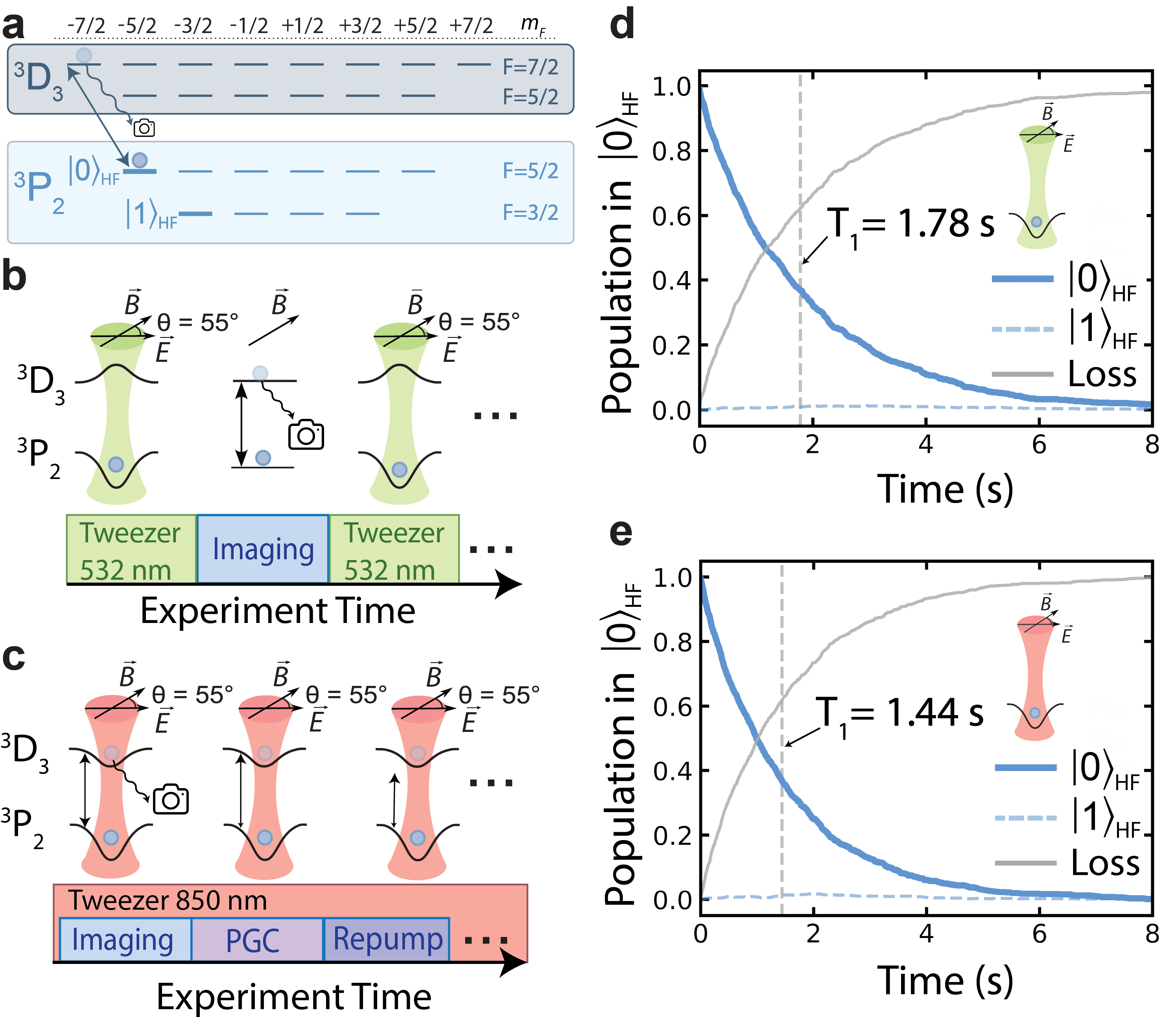}
    \caption{\textbf{Optical trapping and imaging in the $^3\mathrm{P}_2$ manifold.} \textbf{a.} The readout of HF qubits is achieved by selectively imaging $\ket{0}_{\mathrm{HF}}$ through $\ket{^3\mathrm{D}_3,\, F = 7/2,\, m_F = -7/2}$ at $1798$ nm. \textbf{b.} In 532 nm tweezers, the $(6s5d)~\mathrm{^3D_3}$ manifold is anti-trapped, requiring a pulsed imaging protocol to avoid atom heating \cite{zhang2025leveragingerasureerrorslogical}. \textbf{c.} Additionally, we investigate trapping in $850$ nm tweezers, where the $\mathrm{^3D_3}$ manifold is trapped. The imaging process can be interleaved with polarization gradient cooling (PGC) to compensate for heating due to the differential trap depth between $^3\mathrm{P}_2$ and $\mathrm{^3D_3}$. \textbf{d-e.} We estimate the coherence time of HF qubits (initialized in $\ket{0}_{\mathrm{HF}}$) to be $T_1 = 1.78~\mathrm{s}$ in 532 nm tweezers and $T_1 = 1.44~\mathrm{s}$ in $850$ nm tweezers (details in Appendix~\ref{appendix:coherence times}).}
    \label{fig:fig2}
\end{figure}

\begin{figure*}[!ht]
\centering
\includegraphics[width=1\linewidth]{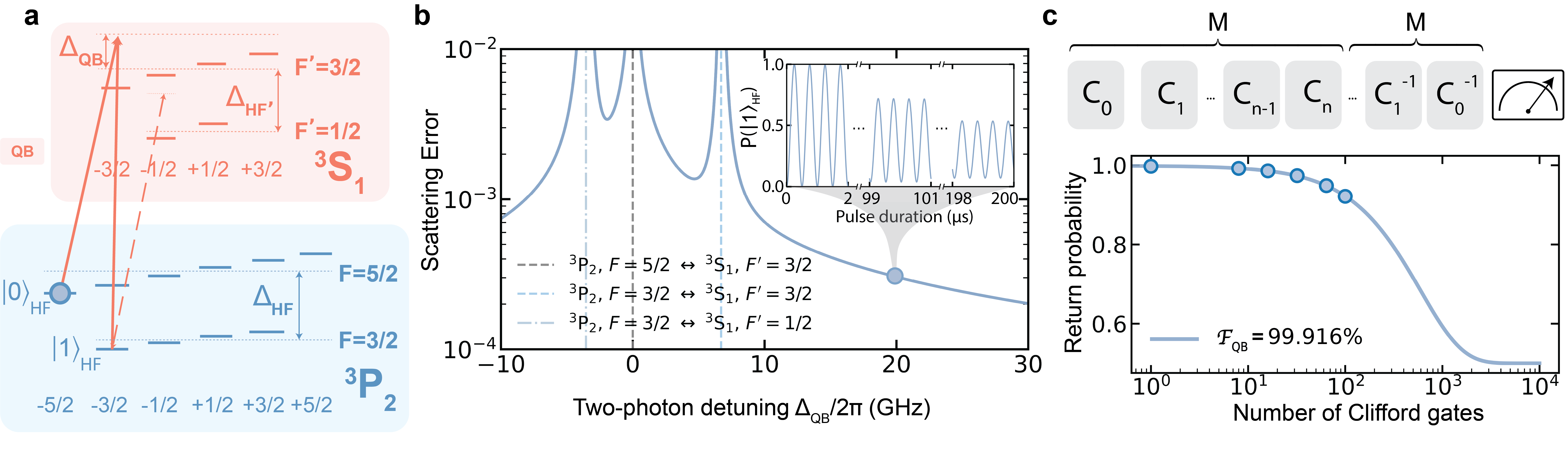}
\caption{\textbf{Single-qubit operation of $^3\mathrm{P}_2$ hyperfine qubits.} \textbf{a.} Level structure and Raman coupling scheme for implementing single-qubit operations within the $^3\mathrm{P}_2$ manifold. The qubit states $\ket{0}_{\mathrm{HF}}$ and $\ket{1}_{\mathrm{HF}}$ are coupled via the intermediate $\ket{^3\mathrm{S}_1, F^{\prime} = 3/2, m_{F^{\prime}} = -3/2}$ state using a two-photon Raman transition. \textbf{b.} Scattering error as a function of the two-photon detuning $\Delta_{\mathrm{QB}}$ with fixed Raman beam powers. To calculate the error, we include Raman and Rayleigh scattering, differential light shifts, and Doppler effects (See Appendix~\ref{appendix:scattering error}). The inset shows the corresponding Rabi oscillations for $\Delta_{\mathrm{QB}} = 2\pi \times 20~\mathrm{GHz}$ and $\Omega_{\mathrm{QB}} = 2\pi \times2~\mathrm{MHz}$. We select an operating point that balances scattering error and coherent coupling strength. \textbf{c.} Simulated randomized benchmarking of single-qubit gates \cite{Knill2008RandomizedBenchmark}. The return probability is shown as a function of the number of Clifford gates. Fitting to a fixed-asymptote decay model yields a single-qubit gate fidelity of $\mathcal{F}_{\mathrm{QB}} = 99.916\%$ (See Appendix~\ref{appendix:operation benchmarking}).} 
\label{fig:fig3}
\end{figure*}

Our encoding utilizes two metastable qubit subspaces in $^{171}$Yb. The $^3\mathrm{P}_0$, $F=1/2$ manifold encodes a purely nuclear spin (NS) qubit in the $m_F$ Zeeman levels, which we denote as $\ket{0}_{\mathrm{NS}} \equiv \ket{^3\mathrm{P}_0,\, F = 1/2,\, m_F = -1/2}$ and $\ket{1}_{\mathrm{NS}} \equiv \ket{^3\mathrm{P}_0,\, F = 1/2,\, m_F = +1/2}$. The $^3\mathrm{P}_2$ $F=3/2$ manifold encodes a hyperfine (HF) qubit in the Zeeman stretched states of the  $F=3/2$ and $F=5/2$ hyperfine levels, which we define as $\ket{0}_{\mathrm{HF}} \equiv \ket{^3\mathrm{P}_2,\, F = 5/2,\, m_F = -5/2}$ and $\ket{1}_{\mathrm{HF}} \equiv \ket{^3\mathrm{P}_2,\, F = 3/2,\, m_F = -3/2}$. The unique characteristics of these encodings inform the qubits' architectural roles.  Due to its nuclear-spin character, the NS qubit provides a coherent computational subspace and a promising storage layer for quantum information. High fidelity single and two-qubit gate operations with NS qubits have already been demonstrated in neutral-atom arrays ~\cite{Jenkins2022Ybqubit,Ma2023YbMidcircuitErasure,Pepper2025MBQT,senoo2025highfidelityentanglementcoherentmultiqubit,MaAlex2022YbUniversalGate}, establishing this encoding as a robust quantum processing resource. The large hyperfine splitting of the HF qubit enables fast single-qubit gate operations, making it well-suited as the primary QEC workspace in our dual-metastable architecture. Combining these qubit encodings realizes a single-species neutral-atom processor that supports long-coherence data qubits and fast mid-circuit syndrome extraction.

Direct readout through state-selective imaging is a key advantage of our HF qubit encoding that is not available in the NS qubit. The closed, cycling transition, $\ket{^3\mathrm{P}_2,\, F=5/2,\, m_F = -5/2} \leftrightarrow \ket{^3\mathrm{D}_3,\, F=7/2,\, m_F = -7/2}$, provides direct imaging of the $\ket{0}_{\mathrm{HF}}$ state at $1798$ nm (where shortwave infrared (SWIR) InGaAs cameras are available for imaging) with a scattering rate of $\Gamma_{\mathrm{read}}=2\pi\times 295$ kHz~\cite{Porsev1999DataYb}. $\ket{0}_{\mathrm{HF}}$ qubits will remain in the $^3\mathrm{P}_2,\, F=5/2$ manifold due to selection rules during imaging, thereby enabling ancilla reuse through reinitialization to the $\ket{0}_{\mathrm{HF}}$ via optical pumping from the other $^3\mathrm{P}_2,\, F=5/2,\, m_F$ levels. This state-selective fluorescence scheme provides a high-contrast readout without disturbing the NS qubits in the computational subspace. These features accelerate syndrome extraction, making the $^3\mathrm{P}_2$ manifold a valuable resource for deep circuit operation. 

Our scheme utilizes a large set of $^{171}\mathrm{Yb}$ electronic states, making it essential to identify tweezer wavelengths that can simultaneously confine all relevant manifolds. We use a master-equation approach to calculate the total light shifts of relevant states across a range of tweezer wavelengths (calculations and polarizability plots are found in Appendix~\ref{appendix:master equation}). We consider only red-detuned tweezers where atoms stay trapped in the high intensity region of tweezers and find that $532~\mathrm{nm}$ trapping light provides attractive potentials for $^1\mathrm{S}_0$, $^3\mathrm{P}_1$, $^3\mathrm{P}_0$, and $^3\mathrm{P}_2$ states, making it a strong candidate for initial tweezer loading. However, $532~\mathrm{nm}$ produces an anti-trapping potential for the $^3\mathrm{D}_3$ state, which is the primary imaging transition channel for the $^3\mathrm{P}_2$ manifold. To mitigate atomic heating during readout of the $\ket{0}_{\mathrm{HF}}$ state, it is necessary to stroboscopically pulse the $532$ nm tweezer to eliminate the anti-trapping potential while the atom is excited to the $\ket{^3\mathrm{D}_3,\, F=7/2,\, m_F = -7/2}$ state (Fig.~\ref{fig:fig2}a) ~\cite{zhang2025leveragingerasureerrorslogical}.

As an alternative, we investigate an additional tweezer trapping wavelength at $850~\mathrm{nm}$. Here, all of the relevant states, including $^3\mathrm{D}_3$, experience attractive trapping potentials. This tweezer wavelength introduces a differential light shift between $^3\mathrm{P}_2$ and $^3\mathrm{D}_3$, but polarization gradient cooling (PGC) can be utilized between imaging pulses to counteract heating, as is commonly done for non-magic imaging (Fig.~\ref{fig:fig2}c) \cite{Wineland:92, PhysRevLett.131.083001}. 

We estimate the $T_1$ coherence time of qubit states in $532~$nm and $850~$nm tweezers based on Raman and Rayleigh scattering rates to nearby excited states and simulate the scattering dynamics through quantum trajectories (shown in Fig.~\ref{fig:fig2}b, d). Here we assume an objective numerical aperture $\mathrm{NA}= 0.65$ and trap depths $\mathrm{U}_0 = 30~\mu$K for both tweezers, corresponding to a power per site of $P = 179~\mu\mathrm{W}$ for $532~\mathrm{nm}$ and $P = 260~\mu\mathrm{W}$ for $850~\mathrm{nm}$, respectively. We assume an atomic temperature $T = 3~\mu$K, which is achievable through gray-molasses and Raman sideband cooling \cite{Lis2023MidcircuitOMG}. To estimate $T_2^{\ast}$, we use $T_2^{\ast} = \sqrt{e^{2/3}-1}\times2 \hbar/(\eta k_{\mathrm{B}}T)$, where $\eta$ is the ratio of differential light shift between the hyperfine qubit states and the electronic ground state light shift \cite{Kuhr2005DephasingTime, Manetsch2025Cs6100array, MaAlex2022YbUniversalGate}. An angle of $\theta=55^{\circ}$ between the tweezer polarization and the quantization axis, set by an external magnetic field, ensures that both HF qubit states experience the same trap depth (See Appendix~\ref{appendix:light shifts} for details). For $532$ nm tweezers we find that $T_1 = 1.78$ s and $T_2^{\ast} = 467.22$ ms, and in $850$ nm tweezers $T_1 = 1.44$ s and $T_2^{\ast} = 82.73$ ms. These coherence values show that both tweezer wavelengths support hundreds of QEC cycles before qubits decohere \cite{Beugnon2007TransferQubit, Levine2022DispersiveHyperfineQubit, Xia2015RBSingleQubit, Guo2020CoherenceTimes}.

\section{Universal Gate Operations}\label{section:universal gate set}

\begin{figure*}[ht!]
  \centering
\includegraphics[width=1\linewidth]{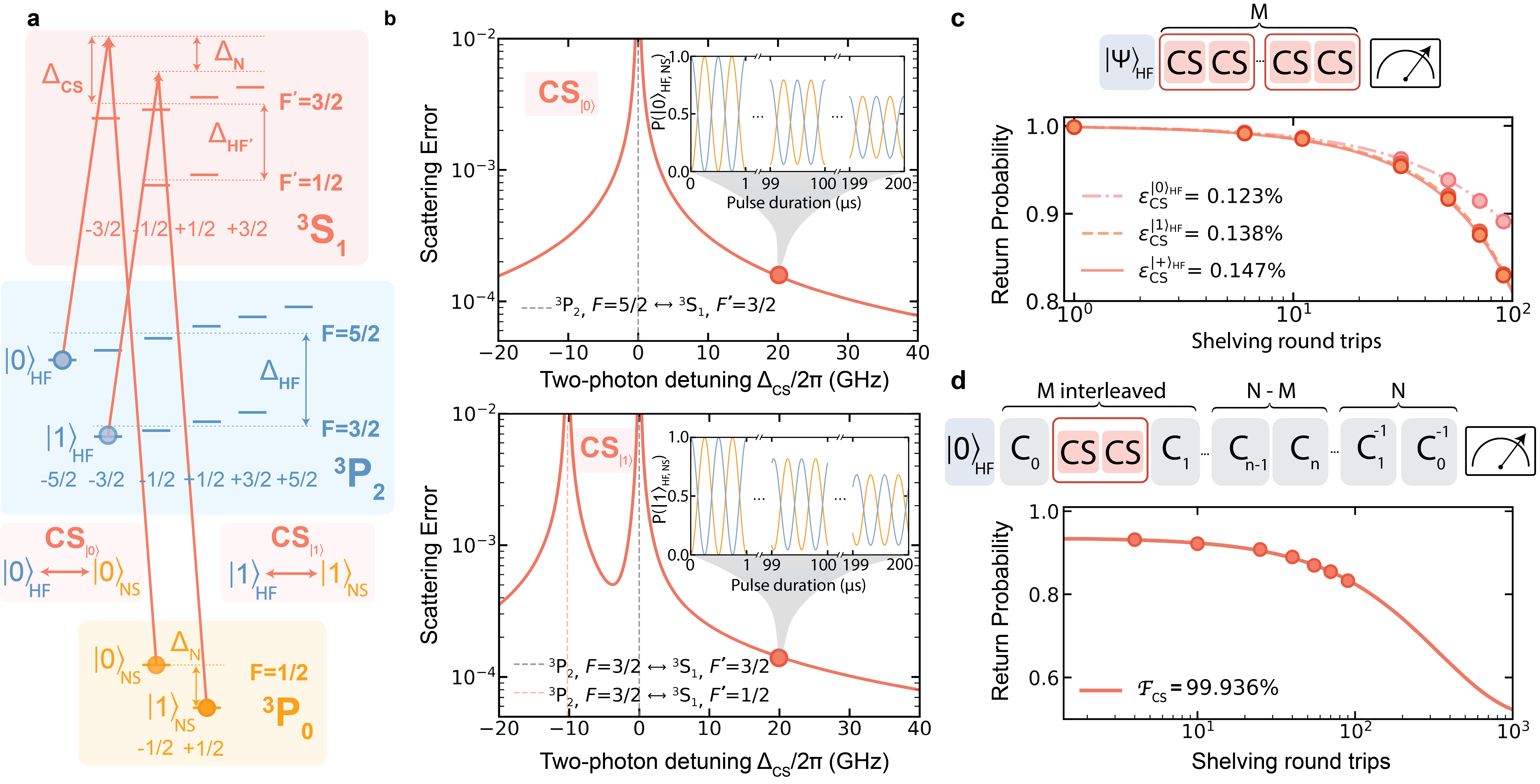}
  \caption{\textbf{Coherent shelving between $^3\mathrm{P}_2$ and $^3\mathrm{P}_0$ manifolds.} \textbf{a.} Level structure and Raman coupling scheme used for the coherent shelving (CS) process. The two Raman pathways, $\mathrm{CS}_{\ket{0}}$ and $\mathrm{CS}_{\ket{1}}$, are modeled using a pair of 770 nm beams detuned by $\Delta_{\mathrm{\mathrm{HF}}} = 2\pi \times 6.7$ GHz, with an additional nuclear-spin splitting $\Delta_N=2\pi\times 11.5$ kHz, and a single $649$ nm beam under a magnetic field $B = 10~\mathrm{G}$. These pathways coherently couple $\ket{0}_{\mathrm{HF}} \leftrightarrow\ket{0}_{\mathrm{NS}}$ and $\ket{1}_{\mathrm{HF}} \leftrightarrow\ket{1}_{\mathrm{NS}}$ via the $^3\mathrm{S}_1$ manifold. \textbf{b.} The scattering error as a function of the two-photon detuning $\Delta_{\mathrm{CS}}$ is computed for a fixed Raman beam power, which yields the Rabi frequency $\Omega_{\mathrm{CS}} = \Omega_{\mathrm{CS}}^{\ket{0}} = \Omega_{\mathrm{CS}}^{\ket{1}} = 2\pi \times 2$ MHz at $\Delta_{\mathrm{CS}}=2\pi\times 20$ GHz (Detailed in Appendix~\ref{appendix:scattering error}). Insets show the corresponding Rabi oscillations, illustrating coherent population transfer. \textbf{c.} Simulated benchmarking of the coherent shelving process. The return probability is monitored as a function of the number of repeated shelving round trips, where each round trip consists of two $\mathrm{CS}(\pi)$ pulses (indicated by red boxes). The resulting decay is fitted to an exponential function to extract the round-trip shelving infidelity $\{\varepsilon_{\mathrm{CS}}^{\ket{0}_{\mathrm{\mathrm{HF}}}},\, \varepsilon_{\mathrm{CS}}^{\ket{1}_{\mathrm{\mathrm{HF}}}},\, \varepsilon_{\mathrm{CS}}^{\ket{+}_{\mathrm{\mathrm{HF}}}}\}$ for three initializing states $\{\ket{0}_{\mathrm{\mathrm{HF}}},\,\ket{1}_{\mathrm{\mathrm{HF}}},\,\ket{+}_{\mathrm{\mathrm{HF}}}\}$. \textbf{d.} Practical implementation of the CS process is modeled using interleaved randomized benchmarking, where $M$ shelving operations are inserted into a fixed-length Clifford sequence ($N = 100$) \cite{Magesan2012InterleavedRB}. The extracted shelving fidelity is $F_{\mathrm{CS}} = 99.936\%$ (Detailed in Appendix~\ref{appendix:operation benchmarking}).} 
  \label{fig:fig4}
\end{figure*}

We now describe the gate primitives required for the HF qubit encoding within our dual-manifold architecture. In addition to the single- and two-qubit gates necessary to implement arithmetic operations and stabilizer-based QEC protocols \cite{Fowler2012SurfaceCode, Xu2024qLDPC}, this architecture requires coherent transfer between the arithmetic and QEC qubit subspaces. Below, we introduce schemes to realize X, Z, and $\mathcal{CZ}$ gates on HF qubits, along with coherent shelving operations that transfer quantum states between the NS and HF qubit subspaces. 

\subsubsection{Single-qubit Operations}
The large hyperfine splitting of the $^3\mathrm{P}_2$ manifold makes the HF qubit a natural platform for fast local operations in our architecture. We propose an optical Raman scheme that utilizes the $^3\mathrm{S}_1$ manifold as the intermediate transition between $\ket{1}_{\mathrm{HF}}$ and $\ket{0}_{\mathrm{HF}}$. We model this operation with a pair of 770 nm beams and include a $B=10~$G magnetic field to provide the frequency selectivity necessary to isolate transitions between the HF qubit states from the other Zeeman sublevels. The selected Raman pathway drives the $\pi$ and $\sigma^+$ transitions from $\ket{1}_{\mathrm{HF}}$ and $\ket{0}_{\mathrm{HF}}$ to $\ket{^3\mathrm{S}_1,\, F^{\prime} = 3/2,\, m_F = -3/2}$ (shown in Fig.~\ref{fig:fig2}a). We model this process using blue-detuned Raman beams that are $\Delta_{\mathrm{QB}} = {2\pi}\times 20$ GHz away from the $\ket{^3\mathrm{S}_1,\,F^{\prime}=3/2}$ manifold in order to minimize the errors from unwanted scattering channels (see Appendix~\ref{appendix:scattering error}). 

We simulate the single-qubit gate fidelity using global randomized benchmarking \cite{Knill2008RandomizedBenchmark, Xia2015RBSingleQubit} (see Appendix \ref{appendix:operation benchmarking}). Specifically, we sample $30$ random circuits from the Clifford group and decompose them into the native gate set $\{\mathrm{R}_{X}(\pi), \mathrm{R}_{X}(\pi/2), \mathrm{R}_{Z}(\pi), \mathrm{R}_{Z}(\pi/2), \mathrm{I}\}$ using  $\texttt{pyGSTi}$~\cite{Nielsen2020-rd}, and vary the circuit depth $M$. The X-rotations are implemented via the Raman process described above, with a fixed Rabi frequency, $\Omega_\mathrm{QB} = {2\pi}\times 2$ MHz. We fit the return probability to a fixed-asymptote decay model (see Appendix~\ref{appendix:master equation}) and obtain an estimated single-qubit gate fidelity of $\mathcal{F}_{\mathrm{\mathrm{QB}}} = 99.916\%$ (see Fig.~\ref{fig:fig3}c). In the present simulation, Z-rotations are implemented via free evolution, allowing the system to accumulate the required phase for the gate. We note that recent experiments have demonstrated fast, local Z-control via light shifts induced by coupling one qubit state to an excited state ~\cite{McKay2017EfficientZgate, Graham2022MultiQubitEntangling, Bluvstein2024QPU, Radnaev2025UnivQPULocalDressing, Wang2025IndividualControlPhase}. Incorporating such techniques would further improve performance, but is not included in the present model. Additionally, decay from $^3\mathrm{S}_1$ to the other $^3\mathrm{P}_J$ manifolds creates a finite probability of population leakage out of the HF computational subspace. We estimate that during a single $\mathrm{R}_{X}(\pi)$ gate, $\sim76\%$ of the leaked population can be converted to erasure errors (detailed in Appendix~\ref{appendix:scattering error}).

\subsubsection{Coherent Shelving}
Our architecture requires coherent shelving (CS) of arbitrary qubit states between the HF and NS qubit subspaces to realize mid-circuit QEC protocols. Our CS scheme utilizes an optical Raman process that transfers atomic population between the ${^3}\mathrm{P}_2$ and ${^3}\mathrm{P}_0$ manifolds via the ${^3}\mathrm{S}_1$ intermediate state. This scheme uses the same $770~$nm Raman beams from the single-qubit gate protocol, but we make both beams $\sigma^+$-polarized. To couple to the ${^3}\mathrm{P}_0$ manifold, the $770~$nm beams are combined with a single $\sigma^-$-polarized $649~$nm beam (see Fig.~\ref{fig:fig4}a). The detuning of all beams is set to $\Delta_{\mathrm{CS}} = 2\pi \times 20~\mathrm{GHz}$ with respect to the $\ket{^3\mathrm{S}_1, F' = 3/2}$  manifold. This configuration couples the qubit states $\ket{0}_{\mathrm{HF}} \leftrightarrow\ket{0}_{\mathrm{NS}}$ and $\ket{1}_{\mathrm{HF}} \leftrightarrow\ket{1}_{\mathrm{NS}}$ via the individual Raman pathways $\mathrm{CS}_{\ket{0}}$ and $\mathrm{CS}_{\ket{1}}$. In practical operations, these pathways must be triggered simultaneously to ensure coherent transfer of arbitrary qubit states between subspaces, which can be achieved with a single 649 nm beam and two 770 nm beams.

A key requirement in this implementation is to identify operating conditions under which the two pathways exhibit identical Raman Rabi frequencies ($\Omega^{\ket{1}}_{\mathrm{CS}} =\Omega^{\ket{0}}_{\mathrm{CS}}  = 2\pi\times 2$ MHz) and accumulate the same phase, which requires compensating light shifts on the qubit states from the Raman beams. To address this, we first simulate the two Raman pathways independently and use the scattering error (shown in Fig.~\ref{fig:fig4}b) to guide the choice of beam powers. We then optimize the $\ket{0}_{\mathrm{HF}}\leftrightarrow\ket{0}_{\mathrm{NS}}$ transition to determine the detuning and intensity of the common 649 nm beam and one of the 770 nm beams. Next, we consider the $\ket{1}_{\mathrm{HF}}\leftrightarrow\ket{1}_{\mathrm{NS}}$ transition, which involves additional off-resonant couplings in ${^3}\mathrm{S}_1$ and calculate the AC Stark shifts induced by the Raman beams (described in Appendix~\ref{appendix:physical model}). We adjust the detuning and intensity of the second 770 nm beam such that both shelving channels exhibit identical transfer dynamics. This realizes a global CS pulse that coherently maps arbitrary superpositions in the HF qubit subspace onto their corresponding states in the NS qubit subspace (the laser parameters used in the simulations are provided in Appendix~\ref{appendix:scattering error}).

To validate of the CS processes, we simulate the population transfer fidelity using a round-trip protocol and monitor the return probability (Fig.~\ref{fig:fig4}c). In our model, we begin with qubits in $\ket{0}_{\mathrm{HF}}$, $\ket{1}_{\mathrm{HF}}$, and $\ket{+}_{\mathrm{HF}}=(\ket{0}_{\mathrm{HF}}+\ket{1}_{\mathrm{HF}}/\sqrt{2})$ states and apply two consecutive CS pulses to transfer the population from the HF qubit subspace to the NS qubit subspace and back. We then fit the return probability for each initial state and extract the following operation errors: $\varepsilon_{\mathrm{CS}}^{\ket{0}_{\mathrm{HF}}}=0.123\%$, $\varepsilon_{\mathrm{CS}}^{\ket{1}_{\mathrm{HF}}}=0.138\%$, and $\varepsilon_{\mathrm{CS}}^{\ket{+}_{\mathrm{HF}}}=0.147\%$ (for details see Appendix~\ref{appendix:operation benchmarking}). This is consistent with the scattering error analysis in Fig.~\ref{fig:fig4}b, where $\mathrm{CS}_{\ket{1}}$ exhibits an additional scattering channel compared to $\mathrm{CS}_{\ket{0}}$. 

In our architecture, CS operations are required for the mid-circuit syndrome-extraction protocol. To characterize the performance of these operations, we simulate randomized benchmarking sequences with CS round-trip pulses and extract the return probability \cite{Magesan2012InterleavedRB, Manetsch2025Cs6100array}. In each sequence, a qubit initialized in  $\ket{0}_{\mathrm{HF}}$ undergoes a fixed-length sequence of $N = 100$ uniformly sampled Clifford gates, with $M$ CS pulse pairs interleaved between them. By varying $M<N$, we fit the decay of the return probability and extract a fidelity of $\mathcal{F}_\mathrm{CS} = 99.936\%$ (see Appendix~\ref{appendix:operation benchmarking} for details). This analysis closely approximates practical operation and confirms that the shelving process preserves coherence for arbitrary qubit states during transfer between manifolds.

\subsubsection{Rydberg-mediated Two-qubit Gates}
\begin{figure}[!t]
    \centering
\includegraphics[width=1\columnwidth]{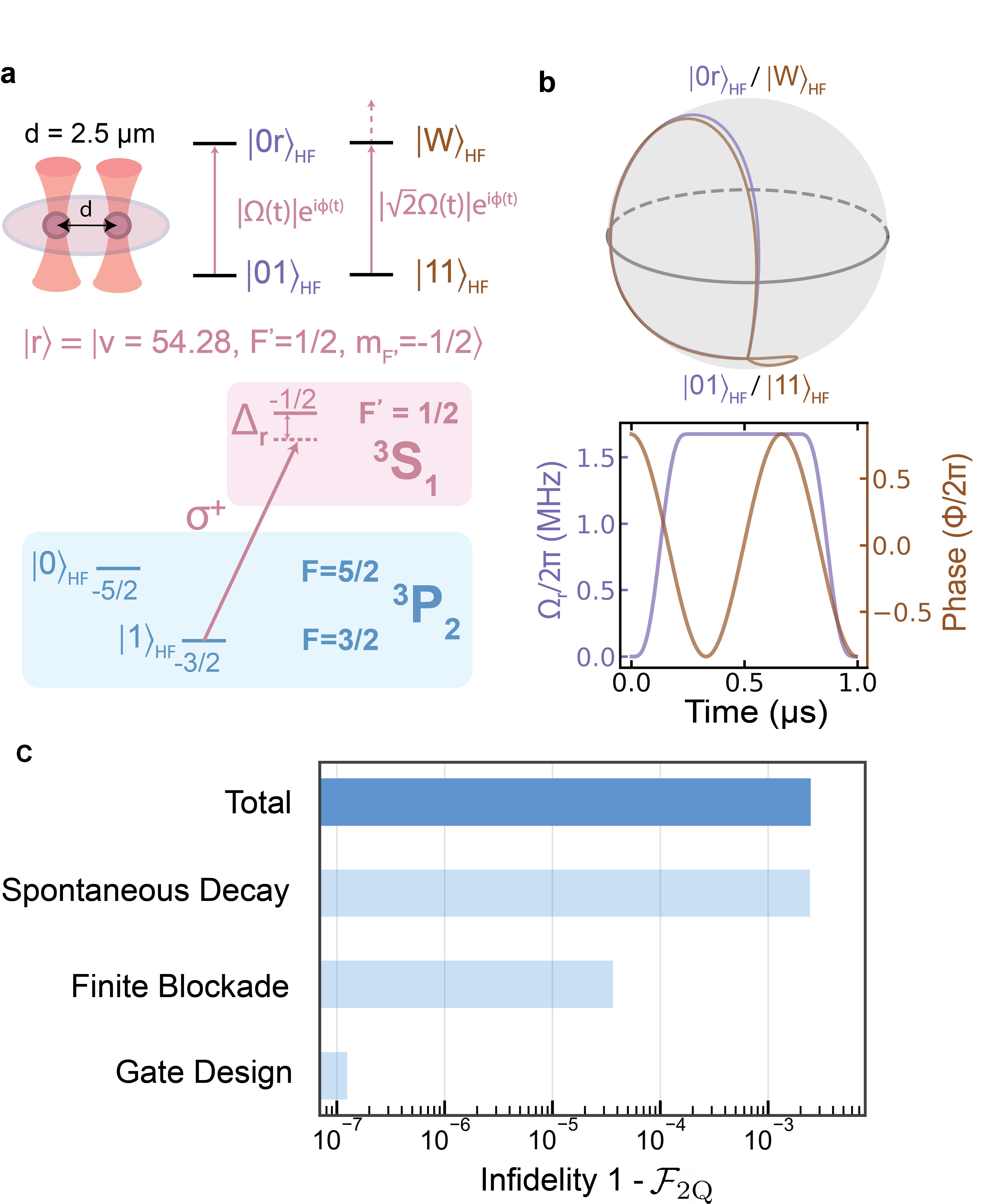}
     \caption{\textbf{Implementing $\textit{CZ}$ gate in the $^3\mathrm{P}_2$ manifold.} \textbf{a.} We selectively couple $\ket{1}_\mathrm{HF} = \ket{^3\mathrm{P}_2, F=3/2, m_F = -3/2}$ to the Rydberg state $\ket{r} \equiv \ket{v=54.28, L=0, F=1/2, m_F=-1/2}$ with a $\sigma^{+}$ polarized beam at $326$ nm and assume the interatomic spacing to be $2.5~\mu\mathrm{m}$. The estimated interaction energy at this spacing is $V = h \times 100~\mathrm{MHz}$ \cite{Pepper2025MBQT}. \textbf{b.} The time-optimal phase gate design (brown solid line) is modeled assuming a smooth-step amplitude function (purple solid line) with peak Rabi frequency $\Omega_{r} = 2\pi \times 1.5 ~\mathrm{MHz}$. The same state and approach is utilized in Refs. \cite{Ma2023YbMidcircuitErasure, zhang2025leveragingerasureerrorslogical}. The resulting time evolution trajectory of the qubit subspace $\{01, 0r\}$ and $\{11, W\}$ is shown on the Bloch sphere, where $\ket{W}_\mathrm{HF} = (\ket{1r}_{\mathrm{HF}} + \ket{r1}_{\mathrm{HF}})/\sqrt{2}$. \textbf{c.} An estimate of the operational error budget, including spontaneous decay, finite blockade, and gate design, gives a two-qubit gate infidelity $1 - \mathcal{F}_\mathrm{2Q} = 0.249\%$.}
    \label{fig:fig5}
\end{figure}

\begin{figure*}[!ht]
    \centering
    \includegraphics[width=\textwidth]{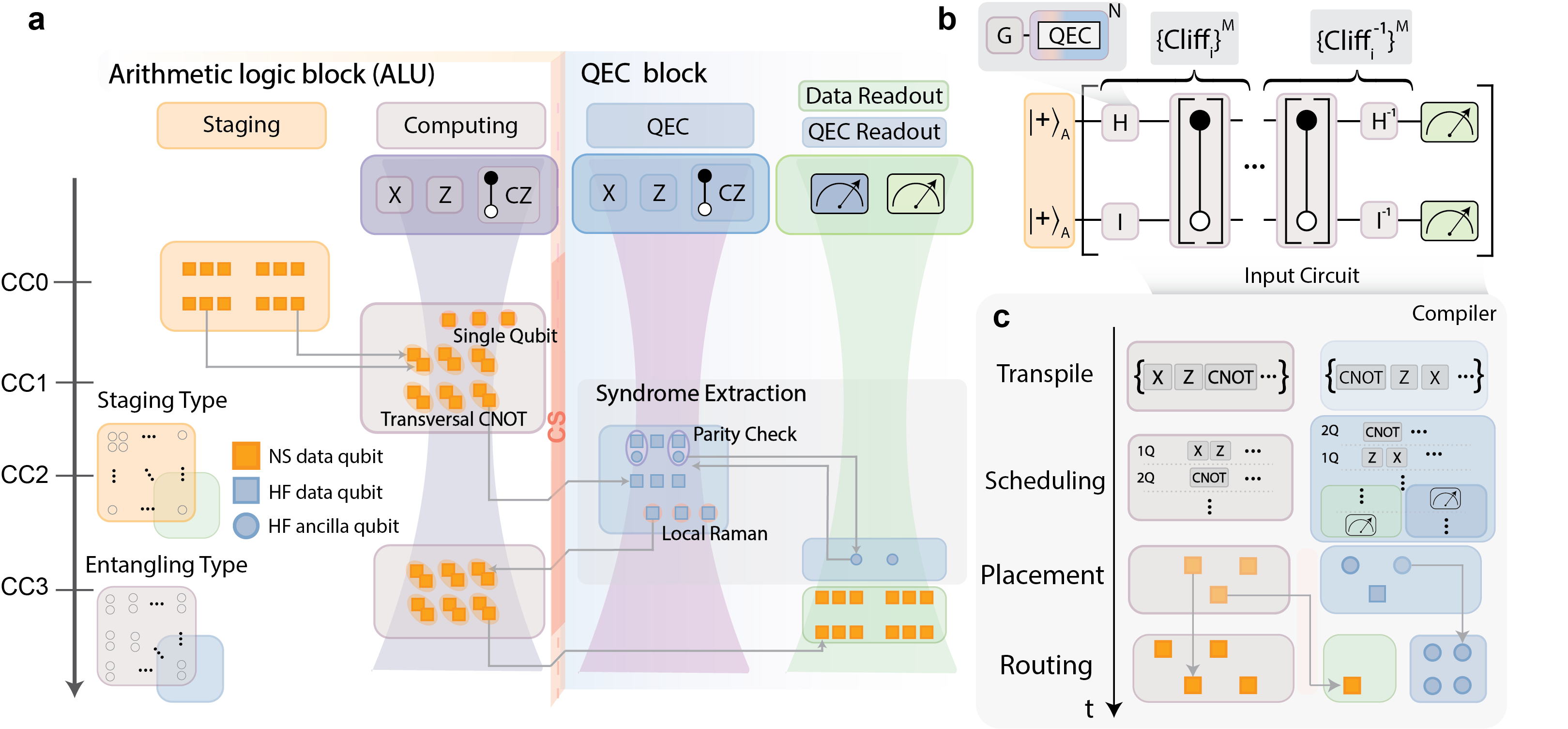}
    \caption{\textbf{Typical mid-circuit cycle and circuit compilation.} \textbf{a.} Data qubits in the arithmetic logic block (ALU) are routed from the staging zone to the computing zone, where we implement logical single-qubit rotations and transversal two-qubit gate operations. Qubits shelved in the QEC block undergo syndrome extraction, in which stabilizer checks are executed using ancilla qubits in the QEC zone. Coherent shelving operations occur when qubits cross the block boundary between the arithmetic and QEC blocks. After syndrome extraction, ancilla qubits are routed to the QEC and readout zones for measurement and reset, after which they can be reused. Data qubits can either remain available for subsequent logical operations or can be shuttled directly to the data readout zone for qubit readout without additional optical shelving. \textbf{b.} Benchmark ansatz used for architectural analysis~\cite{Cain2024CorrelatedDecodingTransversalGate, Ismail2026StarArchitecture}. Two logical qubits evolved through $M$ layers of uniformly sampled logical Clifford gates interleaved with transversal logical CNOT gates. Syndrome extraction of cycles can be performed either mid-circuit or end-circuit, each with $N$ cycles. \textbf{c.} Compilation workflow for mapping circuits onto the architecture for resource analysis. We adapt the compiler design from~\cite{Lin2025NeutralAtomZAC, Stade2025RoutingAware} and make it manifold-aware. The input circuit is transpiled to the native gate set, then scheduled, placed, and routed across zones while accounting for movement and resource constraints.}
    \label{fig:fig6}
\end{figure*}
Metastable states in alkaline-earth atoms support fast, high-fidelity two-qubit gates through single-photon coupling to the Rydberg manifold~\cite{Tsai2025CZLinearResponse, zhang2025leveragingerasureerrorslogical}. To model two-qubit operations in the $^3\mathrm{P}_2$ manifold, we simulate a controlled-phase ($\mathcal{CZ}$) gate protocol following Refs.~\cite{Jandura2022timeoptimaltwothree, Jandura2023OptimizingCZDesign, Ma2023YbMidcircuitErasure}. Our approach selectively couples $\ket{1}_{\mathrm{HF}}$ to the triplet-connected Rydberg state $\ket{r} \equiv \ket{v=54.28, L=0, F=1/2, m_F=-1/2}$ through a circularly polarized $326$ nm beam~\cite{Nakamura2025Rydberg3P2, Pepper2025MBQT}, shown in Fig.~\ref{fig:fig5}a. We choose this Rydberg state due to its interaction strength and isolated pair-energy landscape, previously investigated in \cite{Pepper2025MBQT}.

 We estimate an interaction energy of $V=h\times100~\mathrm{MHz}$ for the Rydberg pair state $\ket{rr}$ at an interatomic separation of $2.5~\mu\mathrm{m}$ using the open source multichannel quantum defect theory (MQDT) software $\texttt{rydcalc}$~\cite{Pepper2025MBQT,Kuroda2025PRA}. The calculation assumes a $10~\mathrm{G}$ magnetic field oriented perpendicular to the interatomic axis with angle $\theta=\pi/2$ and yields $C_6\approx h\times14.4~\mathrm{GHz},\mu\mathrm{m}^6$. We design a time-optimal pulse for a computational subspace spanned by $\ket{0}_{\mathrm{HF}}, \ket{1}_{\mathrm{HF}},$ and $\ket{r}$ following Refs.~\cite{Jandura2022timeoptimaltwothree, Jandura2023OptimizingCZDesign, Ma2023YbMidcircuitErasure,locher2025multiqubitrydberggatesquantum}, and find that the Bell state fidelity is optimized at a detuning of $\Delta_{r}=2\pi \times9~\mathrm{MHz}$ and peak Rabi frequency $\Omega_{\mathrm{r}} = 2\pi \times1.5~\mathrm{MHz}$. These parameters yield a blockade radius of $R_{\mathrm{b}} = (C_6 / \Omega_r)^{1/6} \approx 4.6~\mu m$. We model the pulse amplitude as a smooth step function and track the Bloch sphere trajectories in the two-atom basis under the optimized gate shown in Fig.~\ref{fig:fig5}b. We simulate the error budget (not including Rydberg laser phase noise) under the above conditions and obtain a Bell state infidelity $1 - \mathcal{F}_\mathrm{2Q} = 0.249\%$~\cite{Leseleuc2018RydbergQMCSimulation, Pagano2022ErrorBudgeting, AdamShaw2025SrRandomness, Tsai2025CZLinearResponse}. Additionally, our qubit encoding scheme intentionally omits the ground state $^1\mathrm{S}_0$, leaving it available for future erasure error conversion protocols~\cite{zhang2025leveragingerasureerrorslogical, senoo2025highfidelityentanglementcoherentmultiqubit}.

%% file: pipelined_quantum_processor_architecture.tex
\section{Dual-metastable state zoned quantum processor}\label{section:pipelined processor}
\begin{figure*}[!ht]
    \centering
    \includegraphics[width=\textwidth]{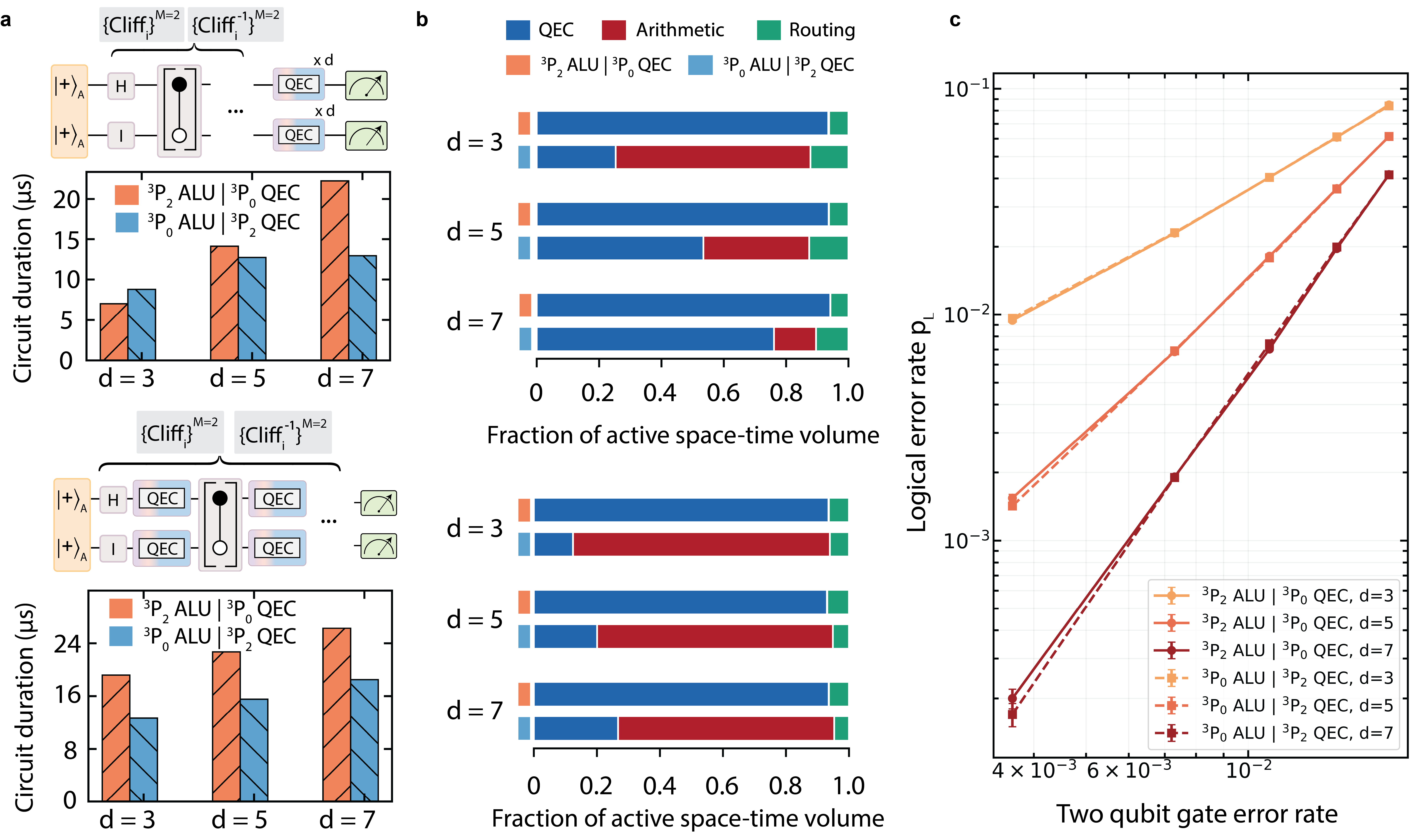}
    \caption{\textbf{Resource estimation and logical-error analysis.} \textbf{a.} Logical benchmark circuits and total execution time for end-circuit and mid-circuit syndrome extraction. In the end-circuit case (upper panel), a number of QEC cycles equal to the code distance $d$ is appended after the randomized benchmark circuit. In the mid-circuit case (lower panel), each logical gate is followed by one QEC cycle, forming a logical gadget. In both cases, we use $M=2$ Clifford gates interleaved with transversal CNOT gates. Bars show the circuit duration for different code distances $d \in \{3,5,7\}$ for the two manifold assignments, with orange denoting $^3\mathrm{P}_2$ as ALU and $^3\mathrm{P}_0$ QEC and blue denoting $^3\mathrm{P}_0$ as ALU and $^3\mathrm{P}_2$ as QEC. \textbf{b.} Fractional active space-time volume of end-circuit (upper panel) and mid-circuit (lower panel) syndrome extraction of the anstaz circuit. Active space-time volume is separated into QEC (dark blue), arithmetic (red), and routing (green). For each manifold assignment and code distance, the plotted fractions are normalized by the total active operation space-time volume, excluding idling. \textbf{c.} Projected logical error rate as a function of the physical two-qubit gate error rate $p_{2Q}$ (for both $^3\mathrm{P}_0$ and $^3\mathrm{P}_2$ qubits). Solid and dashed lines denote the assignments using $^3\mathrm{P}_2$ as the ALU and $^3\mathrm{P}_0$ for QEC, and using $^3\mathrm{P}_0$ as the ALU and $^3\mathrm{P}_2$ for QEC, respectively, while different colors correspond to code distances $d \in \{3,5,7\}$.}
    \label{fig:fig7}
\end{figure*}
We now discuss how the main features of the dual-metastable qubit encoding, including coherent shelving and direct readout, can be leveraged to support mid-circuit protocols in deep circuit regimes within the zoned architecture framework \cite{Bluvstein2022EntanglingTransport}.

Even when using correlated decoding, resource analyses for neutral atom platforms predict $\mathcal{O}(1)$ syndrome extraction rounds per transversal logical gate \cite{Cain2024CorrelatedDecodingTransversalGate, Zhou2025ResourceAnalysis, Ismail2026StarArchitecture}. This indicates that quantum error correction still contributes significantly to both circuit time and resource overhead in near-term machines. Motivated by these considerations, we explore how a dual-metastable encoding scheme can support mid-circuit syndrome measurement in deep circuit regimes. In this section, we focus on implementing Clifford operations and transversal CNOT gates on logical qubits within a surface-code-based framework. 

We organize the architecture of the processor into an arithmetic logic block (ALU) and a QEC block, with qubits separated into ALU and QEC subspaces, as illustrated in Fig.~\ref{fig:fig6}a. Logical operations are performed in the arithmetic block, while repeated syndrome extraction is delegated to the QEC block. A typical mid-circuit processor cycle begins by preparing and arranging data qubits in the staging zone. The data qubits are then transported to the computing zone, where Clifford operations and transversal CNOT gates are executed. After the logical operation, the data qubits are coherently shelved from the arithmetic manifold to the QEC manifold and routed to the QEC block, where they interact with ancilla qubits for stabilizer parity checks. The ancilla qubits are then moved to the readout zone and measured. After the required number of syndrome-extraction cycles, the data qubits are coherently transferred back to the ALU subspace and returned to the arithmetic block for the next logical operation.

The dual-manifold architecture allows operations in the arithmetic block and QEC block to proceed in parallel while maintaining low crosstalk. During syndrome extraction, qubits participating in QEC operations occupy a different metastable manifold from those used for arithmetic operations, so the corresponding control fields are off-resonant from one another. This allows single-qubit operations, shelving, Rydberg excitation, and readout to be distributed across zones and manifolds without enforcing strict spatial or temporal separation. 

Architecture-level performance is directly informed by which qubit manifolds are assigned to the ALU and QEC blocks. The ${^3}\mathrm{P}_0$ and ${^3}\mathrm{P}_2$ manifolds have different operation times and physical error rates, leading to distinct advantages for logical operations and repeated syndrome extraction. We therefore compare the two possible assignments for dual encoding, with $^3\mathrm{P}_0$ used for arithmetic logic and $^3\mathrm{P}_2$ for the QEC in one case, and the roles reversed in the other.

To evaluate each configuration, we use the ansatz circuit shown in Fig.~\ref{fig:fig6}b. The circuit consists of randomly sampled logical Clifford operations interleaved with transversal CNOT gates, followed by the inverse Clifford operation. Syndrome extraction can be applied either at the end of the circuit or inserted after each logical operation (corresponding to an error-corrected logical gadget \cite{Ismail2026StarArchitecture}). This construction probes how repeated logical operations, transversal entangling gates, and mid-circuit syndrome extraction contribute to the architecture-level cost (See Fig.~\ref{fig:fig7}).

We compile the ansatz circuit using a modification of zone-aware compilation (ZAC) \cite{Lin2025NeutralAtomZAC, Stade2025RoutingAware}, summarized in Fig.~\ref{fig:fig6}c. The compiler transpiles the input circuit into the native neutral-atom gate set, schedules single- and two-qubit operations across available zones, places atoms according to zone capacity and reuse constraints, and routes atoms between zones. In addition to the standard zone-aware workflow, we introduce a mapping which specifies the manifold assigned to each zone. This allows the compiler to account for resource analysis such as spatial and temporal utilization of the architecture.

To describe the architecture-level impact of manifold assignments, we use circuit duration and space-time volume for resource analysis. Here, circuit duration is the total time required to implement the ansatz circuits in our architecture, and space-time volume provides a combined metric for the number of qubits used and operation time in each zone (details in Appendix \ref{appendix:architecture_benchmark}). We compare the durations of two benchmarking ansatz circuits which implement end-circuit and mid-circuit syndrome extraction, using the two manifold assignments (Fig.~\ref{fig:fig7}a). For the benchmark circuits considered here, assigning $^3\mathrm{P}_2$ to the QEC block and $^3\mathrm{P}_0$ to the ALU block leads to shorter circuit durations as the code distance increases, owing to faster operation speed in $^3\mathrm{P}_2$ and the ability to reuse ancilla qubits after each syndrome-extraction cycle through non-destructive readout. 

The space-time volume metric allows us to visualize the resource overhead associated with different architecture blocks during ansatz circuit implementation. We further decompose the active space-time volume (excluding idling) into QEC, ALU, and routing for both syndrome-extraction schedules and manifold assignments. The plotted fractions (Fig. \ref{fig:fig7}b) can be interpreted as the distribution of active processor resources. Assigning $^3\mathrm{P}_2$ to the QEC block reduces the relative overhead associated with repeated syndrome extraction and yields a more balanced distribution of active operations. This is particularly apparent in the mid-circuit syndrome-extraction schedule, which suggests an opportunity for parallel execution between the ALU and QEC blocks. As seen in Fig.~\ref{fig:fig7}a, the trend becomes more pronounced at larger code distances, where the cost of syndrome extraction grows, and the manifold assigned to the QEC block has a stronger effect on the overall circuit duration. We therefore choose to assign $^3\mathrm{P}_2$ as the QEC block and $^3\mathrm{P}_0$ as arithmetic logic for the dual-manifold architecture processor design.

We evaluate the logical-level performance of each assignment by constructing physical-level noise models from the operational fidelities presented in Sec.~\ref{section:universal gate set} and sampling the resulting stabilizer circuits using \texttt{Stim} \cite{Bluvstein2024QPU, Rodriguez2025MagicStateDissalation, Sahay2023ErasureMakeHighTresholdCodes, Perrin2025QECAtomLoss, Kobayashi2026ErasureNeutralAtom, Ismail2026StarArchitecture,  gidney2021stim}. The sampled syndromes are decoded with \texttt{PyMatching} using a minimum-weight perfect-matching decoder \cite{pymatchingv2}. Fig.~\ref{fig:fig7}c shows the projected logical error rate as a function of the two-qubit gate error rate in the $^3\mathrm{P}_2$ manifold for several code distances. These results provide insight on how architecture-level scheduling and physical operation errors influence the logical error rate of the dual-metastable architecture.

%% file: conclusion.tex
\section{Discussion}\label{section:conclusion}
In this work, we analyze the potential of a dual-metastable-state qubit encoding in $^{171}\mathrm{Yb}$ for processing architectures that support deep quantum circuits requiring mid-circuit operations and repeated syndrome measurements. Specifically, we focus on the operational capabilities of the $^3\mathrm{P}_2$ manifold, whose hyperfine structure offers fast optical Raman single-qubit gates, direct readout with ancilla reuse, and Rydberg-mediated entangling gates. Qubits encoded in the $^3\mathrm{P}_2$ manifold provide fast control while maintaining a simulated $T_2^* = 82~\mathrm{ms}$ in an $850~$nm tweezer \cite{Levine2022DispersiveHyperfineQubit, Manetsch2025Cs6100array, miles2026qubitsyndromemeasurementshigh}. 
We simulate single-qubit gates and coherent transfer to the $^3\mathrm{P}_0$ manifold with sub-microsecond operation times and fidelities greater than $99.9\%$. We expect further improvements in gate fidelities using quantum optimal control techniques \cite{Jandura2023OptimizingCZDesign}.

Using a framework that combines compiler and error modeling, we compare architecture designs with $^3\mathrm{P}_2$ and $^3\mathrm{P}_0$ qubits and find that utilizing the operation speed of $^3\mathrm{P}_2$ qubits for error correction enables shorter circuit duration and lower resource overhead, particularly in the context of mid-circuit error correction. Our encoding scheme currently does not utilize the ground state$,{^1}\mathrm{S}_0$, which enables the conversion of leakage errors into erasures for both $^3\mathrm{P}_2$ and $^3\mathrm{P}_0$, which can result in higher error thresholds for surface codes \cite{Wu2022ErasureConversionAEA, Ma2023YbMidcircuitErasure, zhang2025leveragingerasureerrorslogical}.  We envision using depumping cycles to properly isolate erasures for each qubit manifold, and expect this to further reduce the resource overhead for error correction in our architecture.

Our dual-manifold encoding offers several resource utilization optimizations. It allows strategic logical qubit layout that can minimize shuttling when interleaving logical operations and mid-circuit readout \cite{Viszlai2025interleaved}. The separation of subspaces supports the temporal overlap of arithmetic and QEC functions through non-blocking operations, allowing different stages of error correction to proceed in parallel across the processor ~\cite{Ramamoorthy1977ClasssicalPipelining, Patomaki2024PipelinedQPUSiliconQubit}. With coherent shelving between manifolds, the encoding allows an operating circuit to dynamically reconfigure the ratio of qubits in each manifold depending on the hardware needs of a process. This encourages a tailored circuit layout that maximizes resource utilization based on the target algorithm and type of QEC code~\cite{Wang2024FPQA, LSQCA, FLASQ}. In future work, we will investigate how these features can improve processor throughput.

Additional advantages of this encoding include support for hierarchical memory protocols using the ultra-long coherence of qubits in the $^3\mathrm{P}_0$ manifold ~\cite{thaker2006quantummemoryhierarchiesefficient, Giovannetti2008QRAM, cesa2025fasterrorcorrectablequantumram, Stein2023HeterogeneousArchitecture, yang2026spacetimeefficienthardwarecompatiblecomplexquantum, Xu2024qLDPC}, and the capability to perform magic state distillation, owing to the operation speeds available in the $^3\mathrm{P}_2$ manifold ~\cite{Rodriguez2025MagicStateDissalation, Kitaev2005MagicStateDistillation}.

The strengths of our dual-manifold encoding establish the metastable manifolds in $^{171}\mathrm{Yb}$ as a versatile platform for practical quantum processing, and open several areas of investigation for optimizing neutral atom-based architectures.

%% file: acknowledgement.tex
\begin{acknowledgements}
 We acknowledge helpful conversations with Shankari Rajagopal, Gokul Ravi, and Jeff Thompson. We thank Shankari Rajagopal and Gokul Ravi for their careful reading of the manuscript. This work was supported by the Army Research Office (Award No. W911NF-25-1-0146) and the Air Force Office of Scientific Research (Award No. FA9550-24-1-0662). EB acknowledges support from the National Science Foundation through the Graduate Research Fellowship Program. The authors declare no competing financial interests.
\end{acknowledgements}

%% file: appendix.tex
\appendix
\section{Physical Level Modeling Framework}\label{appendix:physical model}
The backend of our physical-level simulation consists of two components. The internal state evolution is described by a master-equation approach, incorporating optical pumping, universal gate operations, and randomized benchmarking sequences. The motional dynamics are treated using semiclassical Monte Carlo trajectories, from which we extract the survival probability of atoms in the tweezers under given initial conditions, like tweezer depth and atom temperature.

\subsection{Hyperfine Atomic Master Equation}\label{appendix:master equation}
To describe the complex coupling between hyperfine manifolds, particularly for states with larger total angular momentum (e.g., $^3\mathrm{P}_2$, and $^3\mathrm{D_{3}}$), we develop a more general numerical model of the internal dynamics. We use this model primarily to estimate the fidelity of single-qubit operations and the coherent shelving process. We follow Ref. ~\cite{steck2020QuantumOptics}, starting from the master equation in the hyperfine structure,

\begin{multline}\label{eq:HyperfineMasterEquation}
\partial_t \Tilde{\rho} = -\frac{i}{\hbar}[\Tilde{H}_{\mathrm{A}} + \Tilde{H}_{AL},\Tilde{\rho}] \\
+ \sum_{F_e} \Gamma_{F_e} \frac{2F_e+1}{2F_g+1} \sum_q \Bigl(
C_q \Tilde{\rho} C_q^{\dagger} - \tfrac{1}{2} C_q^{\dagger} C_q \Tilde{\rho} - \tfrac{1}{2} \Tilde{\rho} C_q^{\dagger} C_q
\Bigr).
\end{multline}

We can select an appropriate rotating frame along with its dominating drive to find a set of effective detunings $\Delta_{i}$ and obtain the hyperfine atomic Hamiltonian ~\cite{Einwohner1976GraphBasedRotatingWaveApproximation}

\begin{multline}
\Tilde{H}_{\mathrm{A}}/\hbar = \sum_{F_{e}, m_{F_{e}}} (\delta\omega_{F_{e},m_{F_{e}}} - \Delta_{e})
    \ket{J_{e}, F_{e}, m_{F_{e}}}\bra{J_{e}, F_{e}, m_{F_{e}}},
\end{multline}
and the atom-light interaction Hamiltonian $\Tilde{H}_{AL}$

\begin{multline}
\Tilde{H}_{AL}/\hbar = \\
\quad \sum_{F_{g}, m_{F_{g}}, F_{e}, m_{F_{e}}}
    \bigl(\frac{1}{2}\Omega_{F_{g}, m_{F_{g}}, F_{e}, m_{F_{e}}}
    \ket{F_{e}, m_{F_{e}}}\bra{F_{g}, m_{F_{g}}} + \text{ c.c.}\bigr).
\end{multline}
The term $\delta \omega_{F,m_F}$ accounts for the state-dependent energy shifts arising from both magnetic Zeeman shifts and AC Stark shifts induced by the optical tweezer and applied driving fields. The on-resonance single photon Rabi frequency is obtained from the dipole matrix element,

\begin{multline}
    \hbar \Omega(F, m_F, F', m_F') = \bra{F, m_{F}} \textbf{d} \cdot \textbf{E} \ket{F^{\prime}, m_{F}^{\prime}} =\\
    \quad \left(\frac{2 I_i}{\epsilon_0 c} \right)^{1/2}\bra{F\text{ }m_{F}} er_{q} \ket{F^{\prime}\text{ }m_{F}^{\prime}}.
\end{multline}
We solve the time evolution of the system described by Eq.(\ref{eq:HyperfineMasterEquation}) using \texttt{QuTiP}~\cite{LAMBERT20261Qutip5}. With this scheme, we were able to perform numerical simulations of state initialization, single-qubit rotations within each manifold, shelving between the $^3\textrm{P}_0$ and $^3\textrm{P}_2$ manifolds, and randomized benchmarking.

\subsection{AC Stark Shift and Atomic Polarizability}\label{appendix:light shifts}

 \begin{figure*}[t]
     \centering
     \includegraphics[width=\textwidth]{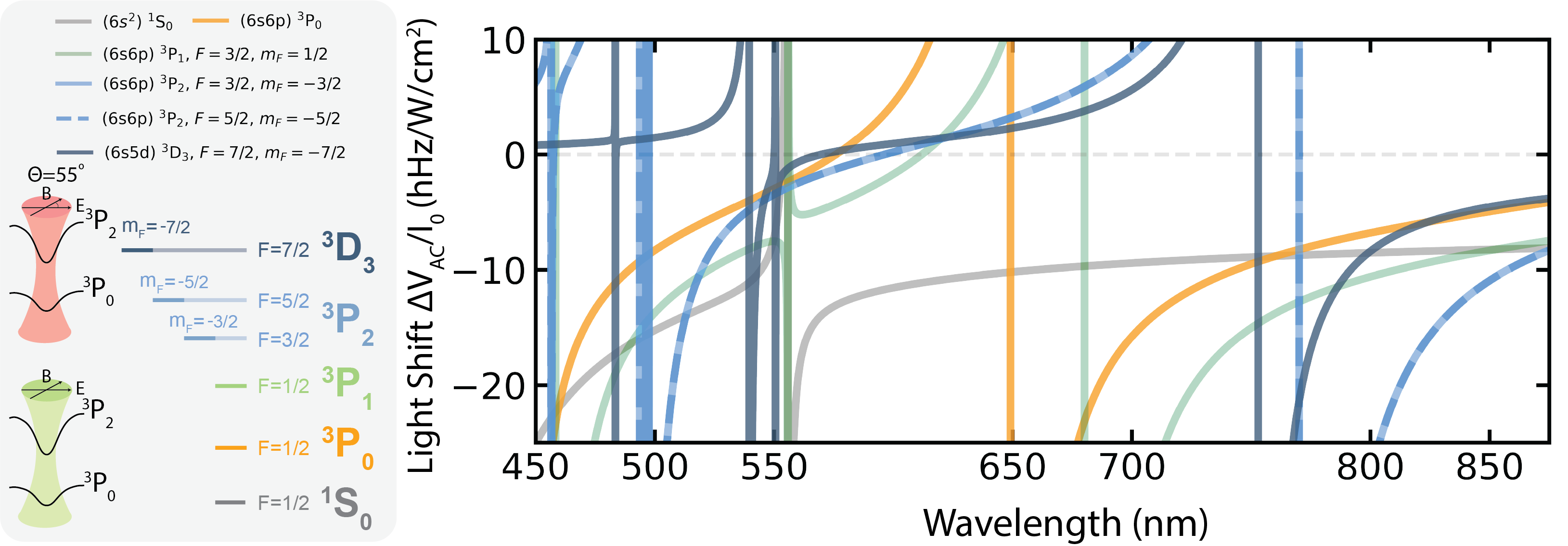}
     \caption{\textbf{Light shifts of relevant states.} Our particular application only considers red-detuned tweezers, trapping a single atom in the intensity maxima. In this case, the tweezer wavelength ranging from $800~$nm to $1000~$nm traps all states of interest. The arithmetic logic block  ($^3\mathrm{P}_2$) use $850~$nm since we only need to trap $^3\mathrm{D}_3$ when we are imaging out of $^3\mathrm{P}_2$. The ground state ($^3\mathrm{P}_2$) has a deeper trap depth compared to the excited state ($^3\mathrm{D}_3$), and we can implement attractive sisyphus cooling.}
     \label{fig:placeholder}
 \end{figure*}

To identify tweezers wavelengths that can trap all necessary atomic states, we calculate the polarizability of each state. The polarizability follows Kramers--Heisenberg polarizability tensor~\cite{Bransden2003AtomBook},
\begin{align}
\alpha_{\mu\nu}(\beta,\omega)
&=
\sum_{\beta'}
\frac{2\omega_{\beta'\beta}}{\hbar}
\frac{
\langle \beta | d_\mu | \beta' \rangle
\langle \beta' | d_\nu | \beta \rangle
}{
\omega_{\beta'\beta}^2 - \omega^2
} .
\end{align}
Here, atomic states $\beta$ are labeled by their term symbols, and
$\hbar\omega_{\beta'\beta}=\hbar(\omega_{\beta'}-\omega_\beta)$
denotes the energy difference between states.
The operator $d_\mu$ represents a Cartesian component of the electric dipole operator
$\mathbf{d}$. The polarizability tensor can be decomposed into scalar, vector, and tensor contributions. The polarizability tensor can be written as
\begin{align}
\alpha_{\mu\nu}
&=
\alpha_s \delta_{\mu\nu}
+
\frac{i\alpha_v}{F}
\epsilon_{\sigma\mu\nu} F_\sigma
\nonumber\\
&\quad
+
\frac{\alpha_t}{2F(2F-1)}
\Big[
3(F_\mu F_\nu + F_\nu F_\mu)
-
2F(F+1)\delta_{\mu\nu}
\Big].
\end{align}
The scalar, vector, and tensor polarizability are given by
\begin{align}
\alpha_s(\beta;\omega)
&=
\sum_{\beta^{\prime}}
\frac{2}{3}
\frac{
\omega_{\beta\beta^{\prime}}
\left|\langle \beta || \mathbf{d} || \beta^{\prime} \rangle\right|^2
}{
\hbar(\omega_{\beta\beta^{\prime}}^2-\omega^2)
} ,
\end{align}

\begin{align}
\alpha_v(\beta;\omega)
&=
\sum_{\beta^{\prime}}
(-1)^{F+F^{\prime}+1}
\sqrt{\frac{6F(2F+1)}{F+1}}
\nonumber\\
&\quad\times
\begin{Bmatrix}
1 & 1 & 1 \\
F & F & F^{\prime}
\end{Bmatrix}
\frac{
\omega_{\beta\beta^{\prime}}
\left|\langle \beta || \mathbf{d} || \beta^{\prime} \rangle\right|^2
}{
\hbar(\omega_{\beta\beta^{\prime}}^2-\omega^2)
} ,
\end{align}

\begin{align}
\alpha_t(\beta;\omega)
&=
\sum_{\beta'}
(-1)^{F+F'}
\sqrt{
\frac{40F(2F+1)(2F-1)}{3(F+1)(2F+3)}
}
\nonumber\\
&\quad\times
\begin{Bmatrix}
1 & 1 & 2 \\
F & F & F^{\prime}
\end{Bmatrix}
\frac{
\omega_{\beta\beta^{\prime}}
\left|\langle \beta || \mathbf{d} || \beta^{\prime} \rangle\right|^2
}{
\hbar(\omega_{\beta\beta^{\prime}}^2-\omega^2)
} .
\end{align}
The quantity
$\langle \beta || \mathbf{d} || \beta^{\prime} \rangle$
is the reduced dipole matrix element (RDME), which can be extracted from either experimentally measured spontaneous decay rates or estimated through branching ratios with measured lifetimes \cite{Porsev1999DataYb, Hohn2026YbTuneOut}. The spontaneous decay rate is related to the RDME via
\begin{align}
\Gamma_{J'\rightarrow J}
&=
\frac{1}{\tau}
=
\frac{\omega_0^3}{3\pi\epsilon_0\hbar c^3}
\frac{2J+1}{2J'+1}
\nonumber\\
&\quad\times
\left|\langle J || e\mathbf{r} || J' \rangle\right|^2 .
\end{align}
The dipole matrix element becomes
\begin{align}
\langle F m_F | e r_q | F' m_F' \rangle
&=
\langle F || e\mathbf{r} || F' \rangle
(-1)^{F'-1+m_F}
\nonumber\\
&\quad\times
\sqrt{2F+1}
\begin{pmatrix}
F' & 1 & F \\
m_F' & q & -m_F
\end{pmatrix}.
\end{align}
The reduced matrix elements are related by
\begin{align}
\langle F || e\mathbf{r} || F' \rangle
&=
\langle J || e\mathbf{r} || J' \rangle
(-1)^{F'+J+1+I}
\nonumber\\
&\quad\times
\sqrt{(2F'+1)(2J+1)}
\begin{Bmatrix}
J & J' & 1 \\
F' & F & I
\end{Bmatrix} .
\end{align}
Assuming a tweezer beam propagating along the $\hat{z}$ axis, the AC Stark shift is \cite{kroeze2026171ybreferencedata}
\begin{align}
V_{\mathrm{ac}}(F,m_F;\omega)
&=
- \frac{I}{2c\epsilon_0}
\Bigg(
\alpha_s(F;\omega)
\nonumber\\
&\quad
+
\alpha_v(F;\omega)\,
\sin(2\gamma)\,
\frac{m_F}{2F}
\nonumber\\
&\quad
+
\alpha_t(F;\omega)
\frac{3m_F^2 - F(F+1)}{2F(2F-1)}
(3\cos^2\theta - 1)
\Bigg) ,
\end{align}
where $\gamma$ parametrizes the ellipticity of the polarization,
$\hat{\epsilon} = \cos\gamma\,\hat{x} + i\sin\gamma\,\hat{y}$,
and $\theta$ is the angle between the electric field and the quantization axis.

\subsection{Optical Pumping Simulation}\label{appendix:optical pumping}
\begin{figure}[!t]
    \centering
    \includegraphics[width=\linewidth]{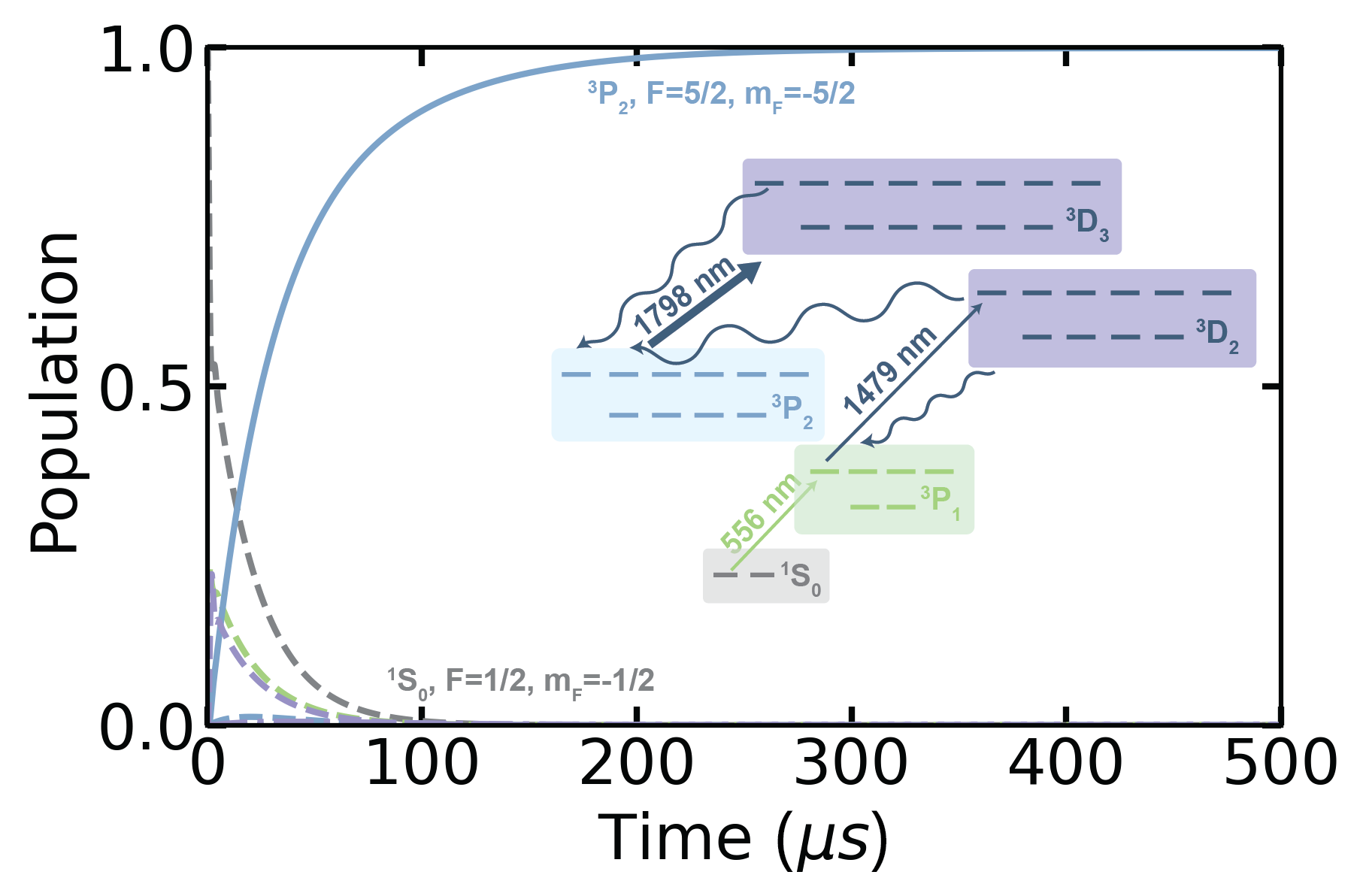}
    \caption{The optical pumping scheme for the initialization of $\ket{0}_{HF}$. Suppose the atoms are initially at $\ket{^1\mathrm{S}_0,\,F=\frac{1}{2},\,m_F=-\frac{1}{2}}$, which can be realized by applying a long 556 nm pulse with $\sigma^-$ polarization. The whole optical pumping to the $^3\mathrm{P}_2$ stretched state can be realized by applying a 556 nm, a 1479 nm, and two 1798 nm pulse simultaneously for 500 $\mu s$, which are all $\sigma^-$ polarization.}
    \label{fig:optical pumping}
\end{figure}

Here we describe the numerical simulations of the proposed optical pumping process. The procedure is modeled using the master-equation solver described in Appendix.~\ref{appendix:master equation}. To initialize atoms to $\ket{0}_{\mathrm{HF}}$ after loading them into tweezer array, the population is first prepared in $\ket{^1\mathrm{S}_0,\,F=1/2,\,m_F = -1/2}$ by depleting $\ket{^1\mathrm{S}_0,\,F=1/2,\, m_F = +1/2}$ through $\ket{^3\mathrm{P}_1,\, F = 3/2,\, m_F = -1/2}$ using a 556 nm beam with $\sigma^-$ polarization. As shown in Fig.~\ref{fig:optical pumping}, atoms in $\ket{^1\mathrm{S}_0,\, F=1/2,\, m_F = -1/2}$ are then excited via a two-photon process to $\ket{^3\mathrm{D}_2,\, F = 5/2,\, m_F = -5/2}$. The decay branching ratios from $^3\mathrm{D}_2$ to $^3\mathrm{P}_2$ and $^3\mathrm{P}_1$ are $12\%$ and $88\%$, respectively. Atoms that decay back to $^3\mathrm{P}_1$ will continue to cycle to $^3\mathrm{D}_2$ and eventually accumulate in the $^3\mathrm{P}_2$ manifold. To prepare all atoms in $\ket{0}_{\mathrm{HF}}$ stretched state, two 1798 nm, $\sigma^-$-polarized beam detuned by $2\pi\times1$ GHz, assisted by a 15 G magnetic field, drive atoms from $\ket{^3\mathrm{P}_2,\, F=5/2,\, 3/2}$ to $^3\mathrm{D}_3$. From $^3\mathrm{D}_3$, subsequent spontaneous decay leads to population accumulation in the dark state $\ket{0}_{\mathrm{HF}}$. Numerical simulations indicate that a population of $99.907\%$ in $\ket{0}_{\mathrm{HF}}$ can be achieved under this optical pumping protocol.

\section{Error Estimation of Raman Transitions}\label{appendix:error estimation}
Physical-level noise in Raman-driven operations, including single-qubit gates in the $^3\mathrm{P}_2$ manifold and coherent shelving, is dominated by photon scattering. We model scattering-induced errors within the qubit subspace using Pauli error channels. Inelastic Raman scattering within the qubit subspace produces population transfer between qubit states and is mapped to a Pauli-$X$ error channel. Elastic Rayleigh scattering instead preserves population but induces dephasing through state-dependent scattering amplitudes, and is mapped to a Pauli-$Z$ channel~\cite{moore2023photon, RamanError, RayleighError}. 

\subsection{Scattering Error}\label{appendix:scattering error}
The scattering contributions are evaluated analytically from the Raman transition amplitudes. Assuming the atom is driven from the initial states $\ket{\beta}$ to the excited states $\ket{k}$ in the $^3\mathrm{S}_1$ manifold by driving field $E_{S\beta}^{}$ with polarization $p \in \{-1, 0, 1\}$, and eventually decay to the final states ${\beta}^\prime$ with a spontaneous decay rate $\Gamma_0^{S\beta^{\prime}}$ with polarization $q \in \{-1, 0, 1\}$. The Raman scattering rate is then given by the summation of the transition amplitudes over all driving channels and decay channels~\cite{moore2023photon}
\begin{equation}
\begin{aligned}
    \Gamma_{Ram}^{}=\sum_{\beta\neq \beta^{\prime};\,q} &\left(\frac{\mu_{S\beta}^{} E_{S\beta}^{}}{2\hbar}\right)^2\Gamma_0^{S\beta^{\prime}}\\
    &\times\left|\sum_{k;\,p}{\frac{\bra{\beta^{\prime}}er_q\ket{k}\bra{k}er_p\ket{\beta}}{\Delta_{k\beta}\mu_{S\beta}^{}\mu_{S\beta^{\prime}}^{}}}\right|^2
\end{aligned}
\end{equation}\label{eq:raman_scattering}
where $\Delta_{e\beta}$ ($\Delta_{e\beta^{\prime}}$) denotes the two-photon detuning, $\mu^{}_{S\beta}$ ($\mu^{}_{S\beta^{\prime}}$) is the spin-orbit coupling dipole-matrix elements between the initial (final) state and the ${^3}\mathrm{S}_1$ excited state, which can be related to the decay rate and transition frequency by $\mu_{mn}^{}=(3\pi\epsilon_0\hbar{c^3}\Gamma_0^{mn}/\omega_{mn}^3)^{1/2}$. Spontaneous Raman scattering occurs when the final state differs from the initial state, while spontaneous Rayleigh scattering corresponds to the case where the initial and final states are identical~\cite{RayleighError}
\begin{equation}
\begin{aligned}
    \Gamma_{Ray}^{} = &\left|\sum_{k;\,p;\,q}\frac{\mu_{S\beta}^{} E_{S\beta}^{}}{2\hbar}\sqrt{\Gamma_0^{S\beta}}{\frac{\bra{\beta}er_p\ket{k}\bra{k}er_p\ket{\beta}}{\Delta_{S\beta}^{}\mu_{S\beta}^2}} \right. \\
    &\left. \qquad - \frac{\mu_{S\beta^{\prime}}^{} E_{S\beta^{\prime}}^{}}{2\hbar}\sqrt{\Gamma_0^{S\beta^{\prime}}}{\frac{\bra{\beta^{\prime}}er_q\ket{k}\bra{k}er_q\ket{\beta^{\prime}}}{\Delta_{S\beta^{\prime}}^{}\mu_{S\beta^{\prime}}^2}}\right|^2
\end{aligned}
\end{equation}\label{eq:rayleigh_scattering}
We can then estimate the scattering error during the Raman transitions via the spontaneous scattering probability over the duration of a $\pi$-pulse
\begin{equation}
    P_{S}^{} = \Gamma_{S}^{}\tau_{\pi}^{} = \frac{\pi\Gamma_{SC}^{}}{\Omega_R^{}}
\end{equation}
where $\Gamma_{S}^{} = (\Gamma_{Ram}^{}+\Gamma_{Ray}^{})/2$. 

Taking into account both spontaneous photon scattering and the requirement of matching Raman Rabi frequencies, we adopt a set of experimentally achievable conditions and use them throughout the simulations presented in this work. After compensating the differential light shifts and Zeeman shifts (see Appendix~\ref{appendix:physical model}) through the detuning of each laser beam, all Raman processes are operated at a common two-photon detuning of $2\pi\times 20$ GHz, which provides a favorable balance between scattering-induced infidelity and experimentally achievable laser power. The corresponding Raman beam configurations used in each simulation are summarized in Table~\ref{tab:beam config}. A more detailed investigation of experimental limitations and optimizations includes laser phase noise, laser linewidth, and larger two-photon detunings which is left for future experimental work.

\begin{table}[ht!]
\centering
\resizebox{\linewidth}{!}{
\begin{tabular}{l c c c c c}
\toprule
\makecell[c]{\textbf{Operation}}~~&
\makecell[c]{\textbf{Wavelength}\\ \textbf{(nm)}}~~&
\makecell[c]{\textbf{Polarization}}~~&
\makecell[c]{\textbf{Ground}\\ \textbf{state}}~~&
\makecell[c]{\textbf{Power}\\ \textbf{(mW)}}~~&
\makecell[c]{\textbf{Beam waist}\\ \textbf{($\mu$m)}} \\
\midrule
\multirow{2}{4em}{QB} 
& 770 & $\sigma^+$ & $\ket{0}_{\mathrm{HF}}$ & 0.24 & 30\\
& 770 & $\pi$ & $\ket{1}_{\mathrm{HF}}$ & 2.38 & 30\\
\hline
\multirow{2}{4em}{$\mathrm{CS}_{\ket{0}}$} 
& 770 & $\sigma^+$ & $\ket{0}_{\mathrm{HF}}$ & 0.24 & 30\\
& 649 & $\sigma^-$ & $\ket{0}_{\mathrm{NS}}$ & 0.71 & 30\\
\hline
\multirow{2}{4em}{$\mathrm{CS}_{\ket{1}}$} 
& 770 & $\sigma^+$ & $\ket{1}_{\mathrm{HF}}$ & 0.58 & 30\\
& 649 & $\sigma^-$ & $\ket{1}_{\mathrm{NS}}$ & 0.71 & 30\\
\hline
\multirow{3}{4em}{$\mathrm{CS}$} 
& 770 & $\sigma^+$ & $\ket{0}_{\mathrm{HF}}$ & 0.24 & 30\\
& 770 & $\sigma^+$ & $\ket{1}_{\mathrm{HF}}$ & 0.59 & 30\\
& 649 & $\sigma^-$ & $\ket{0,1}_{\mathrm{NS}}$ & 0.71 & 30\\
\bottomrule
\end{tabular}
}
\caption{The intensity configurations of Raman beams for the simulations of proposed qubit operations.}
\label{tab:beam config}
\end{table}

\subsection{Qubit Operation Benchmarking}\label{appendix:operation benchmarking}
Under the conditions mentioned in the main text Sec.(\ref{section:universal gate set}), for a single $\mathrm{R}_{X}(\pi)$ gate, we estimate that approximately $55\%$ of the leaked population eventually decays to $^1\mathrm{S}_0$, while about $21\%$ ends in $^3\mathrm{P}_0$. Both cases can be converted into erasure errors by applying a repumping pulse. The remaining $\sim 24\%$ remains within nearby magnetic sublevels of the $^3\mathrm{P}_2$ manifold through Raman scattering. 

Two methods are provided to assess coherent shelving. The first method apply $M$ round trips of shelving for given initial states $\ket{0}_{\mathrm{HF}},\,\ket{1}_{\mathrm{HF}},\,\ket{+}_{\mathrm{HF}}$ states. By fitting the return probability into exponential decay model~\cite{yang2022dualtypeion}
\begin{equation}
    P_r(M)=A(1-\varepsilon_{\mathrm{CS}})^M+(1-A)
\end{equation}
we can extract the round-trip shelving error rate $\varepsilon_{\mathrm{CS}}^{\ket{0}_{\mathrm{HF}}},\,\varepsilon_{\mathrm{CS}}^{\ket{1}_{\mathrm{HF}}},\,\varepsilon_{\mathrm{CS}}^{\ket{+}_{\mathrm{HF}}}$, as shown in Fig.~\ref{fig:fig4}c (upper panel). The lower error rate of $\mathrm{CS}_{\ket{0}}$ arises from its cleaner three-level coupling pathway and the stretched-state character of the transition. In comparison, $\mathrm{CS}_{\ket{1}}$ is more susceptible to additional excited-state scattering pathways and weak off-resonant coupling by the other $770$ nm beam from $\ket{1}_{\mathrm{NS}}$ to $\ket{{^3}\mathrm{P}_2,\,F=5/2,\,m_F=-3/2}$, leading to leakage outside the qubit subspace.

The second method leverages the interleaved randomized benchmarking sequence to extract effective depolarizing channels for the proposed scheme. Our work currently does not include laser intensity and phase noises. For single-qubit operations in the $^3\mathrm{P}_2$ manifold, the depolarizing channel is extracted using randomized benchmarking. We assume atomic temperature $3~\mu\mathrm{K}$ and with counter-propagating Raman beams, we sample 30 circuits from the Clifford group at various depths $M$ using $\texttt{pyGSTi}$ \cite{Nielsen2020-rd}, and decompose each circuit into the native gate set. Here our simulation assume the SPAM free case. We evolve each circuit from a given initial state, measure the return probability $\mathrm{P}_{r}$, and fit it using the zeroth-order model 

\begin{equation}\label{eq:fix2_asympotic_RB}
    P_{r}(M) = A\alpha^M+1/d,
\end{equation}
where $\alpha$ is the depolarizing parameter, $d = 2^{n}$ is the dimension of qubit sub-space, and $r$ is the averaged Clifford error rate $r = (d-1)(1-\alpha)/d$. One can relate the depolarizing error rate $p = 1 - \alpha = 2r$ for single qubit operation ($n = 1$)~\cite{Knill2008RandomizedBenchmark, Magesan2012InterleavedRB, Magesan2012RBTheory}. We then define $r_{\mathrm{QB}}$ as average error rate of the single-qubit gate \cite{Magesan2012InterleavedRB, Magesan2012RBTheory} in $^3\mathrm{P}_2$. This gives an averaged Clifford gate fidelity \cite{Knill2008RandomizedBenchmark} of $F_{\mathrm{QB}} = 1 - r_{\mathrm{QB}} = 99.916\%$, as shown in Fig.~\ref{fig:fig3}c.

For coherent shelving, we extract the depolarizing error rate $p_{\mathrm{CS}}$ using interleaved randomized benchmarking. Specifically, we interleave $M$ round trips of the coherent shelving operation, $\mathrm{CS}(\pi)$, with a fixed total number $N = 100$ of $^3\mathrm{P}_2$ single-qubit operations, using $30$ random circuit samples for each value of $M$. We again fit the return probability using Eq.\ref{eq:fix_asympotic_RB} and obtain depolarizing parameter $\Tilde{\alpha}_{\mathrm{CS}}$ per shelving round-trip. The estimated average error rate per shelving round-trip is $\Tilde{r}_{\mathrm{CS}} = \frac{1}{2} (d - 1)(1 - \Tilde{\alpha}_{\mathrm{CS}}/\alpha_{\mathrm{QB}})$ \cite{Magesan2012InterleavedRB}. The round-trip fidelity of coherent shelving is then as $\Tilde{F}_{\mathrm{CS}} = 1 - \Tilde{r}_{\mathrm{CS}} = 99.936\%$ as shown in Fig.~\ref{fig:fig4}c (lower panel). And we estimate the depolarizing error rate per single path shelving as $p_{\mathrm{CS}} = 1 - \sqrt{\Tilde{\alpha}_{\mathrm{CS}}}$.

\section{Tweezer Coherence Times}\label{appendix:coherence times}
We estimate the $T_{1}$ coherence time through off-resonant scattering from the tweezers. We adapt the result from Ref.\cite{Dorscher2018ScatteringRate} and the general dipole operator to $d_{q} = \Sigma\epsilon_{p} d_{p}$ \cite{Martin2013SrClock}. Assume that we have the quantization axis align along the z-axis, the scattering rate $\Gamma_{\beta \to \beta^{\prime}}$ of an initial state $\ket{\beta}$ to $\ket{\beta^{\prime}}$ resulting from a far-detuned trapping beam with polarization $p \in \{-1, 0, 1\}$ and intensity $I$ is \cite{Martin2013SrClock}

\begin{equation}
\begin{split}
\Gamma_{\beta \to \beta^{\prime}}
=
C \sum_{q=-1}^{+1}
\bigg|
\sum_k \sum_{p=-1}^{+1}
\epsilon_p
\bigg(
&
\frac{
\langle \beta^{\prime}|e r_q|k\rangle
\langle k|e r_p|\beta\rangle
}{
\omega_{k\beta} - \omega
}
\\
&+
\frac{
\langle \beta^{\prime}|e r_p|k\rangle
\langle k|e r_q|\beta\rangle
}{
\omega_{k\beta} + \omega
}
\bigg)
\bigg|^2
\end{split},
\end{equation}
where we iterate through all dipole allowed transitions with intermediate states $\ket{k}$, and $\omega$ is the tweezer frequency and $\omega_{k\beta}$ is the transition frequency between states $\ket{k}$ and $\ket{\beta}$, and 

\begin{equation}
C
=
 \frac{8\pi}{3} \cdot \frac{I \, \omega^3}{(4\pi \epsilon_0)^2 c^4 \hbar^3}
.
\end{equation}
We then take the scattering rates among $\ket{0}, \ket{1}$, and a virtual state representing leakage out of qubit sub-space and simulate the dynamic through quantum Monte-Carlo trajectories \cite{Nowak2008QuantumTrajectory, Clemens2003CollectiveSpontaneousEmission, Daley2014QuantumTrajectory}.

\section{Architecture benchmark and simulation details}
\label{appendix:architecture_benchmark}

The architecture benchmark in Fig.~\ref{fig:fig6} uses an ansatz circuit inspired by interleaved randomized benchmarking \cite{Magesan2012InterleavedRB, Cain2024CorrelatedDecodingTransversalGate, Ismail2026StarArchitecture}. The circuit consists of $M$ uniformly sampled logical Clifford gates, each followed by a transversal CNOT gate. The Clifford gates are sampled from $\{H,I,S\}$, and an inverse Clifford operation is applied at the end so that the ideal circuit returns to a known output state. For circuits with more than two logical qubits, transversal CNOT gates are applied in an alternating brickwork pattern. We compare two syndrome-extraction schedules. In the mid-circuit schedule, $N=1$ syndrome-extraction round is applied after each logical operation. In the end-circuit schedule, syndrome extraction is deferred until the end of the benchmark block, where $N=d$ rounds are applied. The resource-estimation results in Fig.~\ref{fig:fig7}a,b use two logical code patches with circuit depth $M=2$ and compare both syndrome-extraction schedules.

The ansatz circuit is compiled using the manifold-aware extension we developed based on zone-aware compilation \cite{Lin2025NeutralAtomZAC, Stade2025RoutingAware}. The input circuit is first transpiled into the native neutral-atom gate set $\{\mathrm{U}_3,\mathcal{CZ}\}$. The scheduler then organizes single- and two-qubit operations into time layers and assigns them to processor zones subject to zone-capacity constraints. Placement and routing determine the physical sites and atom transport paths needed to implement the scheduled operations. A manifold-aware zone map specifies whether each zone is assigned to the $^3\mathrm{P}_0$ or $^3\mathrm{P}_2$ manifold, allowing the compiler to use the corresponding operation durations when estimating the total circuit duration and space time volume as shown in Fig.~\ref{fig:fig7}a,b. We define the space-time volume reported in Fig.~\ref{fig:fig7}b as
\begin{equation}
    V_{\mathrm{space-time}} = \sum_{\{z, o\}} N_{z, o} t_{z,o} ,
\end{equation}
where the sum runs over architectural zones and operations, $N_{z, o}$ is the number of occupied physical sites or atoms in zone $z$ executing operation $o$, and $t_{z,o}$ is the total time those atoms spend when executing the operation in the assigned zone. The active space-time volume includes all non-idle operations in the circuit. These include gate execution, readout, and reset in the arithmetic and QEC blocks, together with routing and other scheduled control operations.

The circuit-level noise model is constructed from physical error rates extracted from randomized benchmarking, as discussed in the main text and Appendix~\ref{appendix:scattering error}. For $^3\mathrm{P}_0$, we use hardware-derived error rates from Refs.~\cite{zhang2025leveragingerasureerrorslogical, senoo2025highfidelityentanglementcoherentmultiqubit}, as summarized in Table~\ref{tab:combined_noise_model}. For the logical-level results in Fig.~\ref{fig:fig7}c, we assign the $^3\mathrm{P}_2$ manifold to the QEC block and the $^3\mathrm{P}_0$ manifold to the arithmetic block. Noise is injected according to both the location and timing of each operation in the compiled circuit, making the circuit-level noise model manifold-aware. We mainly follow Ref.~\cite{Ismail2026StarArchitecture} for the physical-level noise model. State-preparation errors are applied after reset operations and are modeled as a depolarizing channel. Single-qubit gate errors are applied after each physical single-qubit operation and are modeled as a depolarizing channel. Shuttling errors are applied whenever qubits are moved and are modeled as a biased Pauli channel, followed by a $Y$ channel representing loss during transport. Measurement errors are modeled as bit-flip $X$ errors applied before measurement, which flip the reported measurement outcome. Two-qubit gate errors are applied after each CNOT operation and are modeled as a biased two-qubit Pauli channel \cite{Sahay2023ErasureMakeHighTresholdCodes}. Shelving errors are inserted whenever a qubit is transferred between the arithmetic and QEC zones, corresponding to a change between the $^3\mathrm{P}_0$ and $^3\mathrm{P}_2$ manifolds, and are modeled as a depolarizing channel. We model the decoherence effects from $T_1$ and $T_2^{\ast}$ with the Pauli twirling approximation \cite{Ghosh2012TwirlingDecoherence, Viszlai2025interleaved}.
A single-qubit Pauli channel with error probabilities $\{p_X,p_Y,p_Z\}$ is written as
\begin{equation}
\mathcal{E}_{1\mathrm{Q}}(\rho)
=
(1-p)\rho
+
\sum_{M\in\{X,Y,Z\}} p_M M\rho M^{\dagger},
\end{equation}
where $p=p_X+p_Y+p_Z$. The single-qubit depolarizing channel is the special case in which the error is distributed uniformly over the three non-identity Pauli operators, $p_X=p_Y=p_Z=p/3$.

Similarly, two-qubit gate errors are modeled using a two-qubit Pauli channel,
\begin{equation}
\mathcal{E}_{2\mathrm{Q}}(\rho)
=
(1-p)\rho
+
\sum_{M\in\mathcal{P}}p_{M}
M\rho M^{\dagger},
\end{equation}
where $p=\sum_{M\in\mathcal{P}}p_M$ and
\begin{equation}
\begin{aligned}
\mathcal{P}=\{&
IX,IY,IZ,XI,XX,XY,XZ,YI,\\
&YX,YY,YZ,ZI,ZX,ZY,ZZ\}.
\end{aligned}
\end{equation}
A two-qubit depolarizing channel is the special case in which the total two-qubit error probability is distributed uniformly over these 15 non-identity two-qubit Pauli operators, i.e., $p_M=p/15$ for all $M\in\mathcal{P}$.

Loss and erasure-type errors are not included in the current simulation framework. These errors can be treated separately from Pauli errors because they correspond to detectable leakage or atom loss, rather than an unknown Pauli error acting on qubits \cite{Sahay2023ErasureMakeHighTresholdCodes, Perrin2025QECAtomLoss, Kobayashi2026ErasureNeutralAtom}. We leave a detailed investigation of erasure-aware noise modeling and decoding to future work.

The resulting stabilizer circuits are sampled using Stim \cite{gidney2021stim}. The sampled syndromes are decoded with PyMatching \cite{pymatchingv2} using a minimum-weight perfect-matching decoder. The logical error rate is extracted as a function of the two-qubit gate error rate $\epsilon_{\mathrm{2Q}}$ in the $^3\mathrm{P}_2$ manifold for different surface-code distances.

\begin{table}[ht!]
\centering
\resizebox{\linewidth}{!}{%
\begin{tabular}{l c c c c}
\toprule
\textbf{$^3\mathrm{P}_2$ physical operation}~
& \textbf{Error}~
& \textbf{Duration ($\mu$s)}~
& \textbf{Noise channel}~\\
\midrule
State preparation$^{\dagger}$
& $0.929\%$
& $500$
& depolarizing \\

Single-qubit gate$^{\dagger}$
& $0.084\%$
& $0.25$
& depolarizing \\

Two-qubit$\mathcal{CZ}$ gate $^{\dagger}$
& $0.249\%$
& $0.25$
& Pauli \\

Atom transport$^{\S}$
& $0.5\%$
& $15$
& Pauli, Y-error\\

Coherent shelving$^{\dagger}$
& $0.064\%$
& $0.25$
& depolarizing \\

Qubit Readout$^{\ddagger}$
& $0.1\%$
& $4000$
& X-error \\

Coherence $T_1$$^{\dagger}$
& $1.44~\mathrm{s}$
& $-$
& Pauli Twirling \\

Dephasing $T_2^{\ast}$$^{\dagger}$
& $0.082~\mathrm{s}$
& $-$
& Pauli Twirling \\

\bottomrule
\end{tabular}%
}
\caption{Summary of physical operation errors, durations, and corresponding noise channels for qubits encoded in the $^3\mathrm{P}_2$ manifold. $^{\dagger}$ Values derived from Sec.(\ref{section:qubit encoding and trapping}) and Sec.(\ref{section:universal gate set}). $^{\ddagger}$ Estimated by comparing the achievable scattering rate of the $^3\mathrm{P}_2 \leftrightarrow ^3\mathrm{D}_3$ with $^1\mathrm{S}_0$ with method mentioned in \cite{Wu2022ErasureConversionAEA}.$^{\S}$ As this value is limited by the atomic temperature and trajectory \cite{zhang2025leveragingerasureerrorslogical,Manetsch2025Cs6100array}, we adopt the value reported in Ref.~\cite{Viszlai2025interleaved}
as a representative estimate.}
\label{tab:combined_noise_model}
\end{table}

%% file: reference.bib
@article{Schlosser2001Subpoisson,
	author = {Schlosser, Nicolas and Reymond, Georges and Protsenko, Igor and Grangier, Philippe},
	journal = {Nature},
	number = {6841},
	pages = {1024--1027},
	title = {Sub-poissonian loading of single atoms in a microscopic dipole trap},
	volume = {411},
	year = {2001},
    url = {https://doi.org/10.1038/35082512}
}

@article{Bernien2017AtomArray,
	author = {Bernien, Hannes and Schwartz, Sylvain and Keesling, Alexander and Levine, Harry and Omran, Ahmed and Pichler, Hannes and Choi, Soonwon and Zibrov, Alexander S. and Endres, Manuel and Greiner, Markus and Vuletić, Vladan and Lukin, Mikhail D.},
	journal = {Nature},
	number = {7682},
	pages = {579--584},
	title = {Probing many-body dynamics on a 51-atom quantum simulator},
	volume = {551},
	year = {2017},
    url = {https://doi.org/10.1038/nature24622}
}

@article{Manetsch2025Cs6100array,
	author = {Manetsch, Hannah J. and Nomura, Gyohei and Bataille, Elie and Lv, Xudong and Leung, Kon H. and Endres, Manuel},
	journal = {Nature},
	number = {8088},
	pages = {60--67},
	title = {A tweezer array with 6,100 highly coherent atomic qubits},
	volume = {647},
	year = {2025},
    url = {https://doi.org/10.1038/s41586-025-09641-4}
}

@article{Aaron2026Metasurface,
  author={Holman, Aaron and Xu, Yuan and Sun, Ximo and Wu, Jiahao and Wang, Mingxuan and Zhu, Zezheng and Seo, Bojeong and Yu, Nanfang and Will, Sebastian},
  journal={Nature},
  pages={859–-865},
  title={Trapping of single atoms in metasurface optical tweezer arrays},
  year={2026},
  volume = {649},
  url = {https://doi.org/10.1038/s41586-025-09961-5}
}

@article{Schymik2021Lifetime6000s,
  title = {Single Atoms with 6000-Second Trapping Lifetimes in Optical-Tweezer Arrays at Cryogenic Temperatures},
  author = {Schymik, Kai-Niklas and Pancaldi, Sara and Nogrette, Florence and Barredo, Daniel and Paris, Julien and Browaeys, Antoine and Lahaye, Thierry},
  journal = {Phys. Rev. Appl.},
  volume = {16},
  issue = {3},
  pages = {034013},
  numpages = {8},
  year = {2021},
  month = {Sep},
  publisher = {American Physical Society},
  doi = {10.1103/PhysRevApplied.16.034013},
  url = {https://link.aps.org/doi/10.1103/PhysRevApplied.16.034013}
}

@article{Chiu2025ContineousLoading,
	author = {Chiu, Neng-Chun and Trapp, Elias C. and Guo, Jinen and Abobeih, Mohamed H. and Stewart, Luke M. and Hollerith, Simon and Stroganov, Pavel L. and Kalinowski, Marcin and Geim, Alexandra A. and Evered, Simon J. and Li, Sophie H. and Lyu, Xingjian and Peters, Lisa M. and Bluvstein, Dolev and Wang, Tout T. and Greiner, Markus and Vuleti{\'c}, Vladan and Lukin, Mikhail D.},
	date = {2025/10/01},
	doi = {10.1038/s41586-025-09596-6},
	id = {Chiu2025},
	isbn = {1476-4687},
	journal = {Nature},
	number = {8087},
	pages = {1075--1080},
	title = {Continuous operation of a coherent 3,000-qubit system},
	url = {https://doi.org/10.1038/s41586-025-09596-6},
	volume = {646},
	year = {2025}
    }

@article{li2025fastcontinuouscoherentatom,
  title={Fast, continuous and coherent atom replacement in a neutral atom qubit array},
  author={Li, Yiyi and Bao, Yicheng and Peper, Michael and Li, Chenyuan and Thompson, Jeff D},
  journal={arXiv:2506.15633},
  url = {https://doi.org/10.48550/arXiv.2506.15633},
  year={2025}
}

@article{Rodriguez2025MagicStateDissalation,
	FULLauthor = {Sales Rodriguez, Pedro and Robinson, John M. and Jepsen, Paul Niklas and He, Zhiyang and Duckering, Casey and Zhao, Chen and Wu, Kai-Hsin and Campo, Joseph and Bagnall, Kevin and Kwon, Minho and Karolyshyn, Thomas and Weinberg, Phillip and Cain, Madelyn and Evered, Simon J. and Geim, Alexandra A. and Kalinowski, Marcin and Li, Sophie H. and Manovitz, Tom and Amato-Grill, Jesse and Basham, James I. and Bernstein, Liane and Braverman, Boris and Bylinskii, Alexei and Choukri, Adam and {DeAngelo}, Robert J. and Fang, Fang and Fieweger, Connor and Frederick, Paige and Haines, David and Hamdan, Majd and Hammett, Julian and Hsu, Ning and Hu, Ming-Guang and Huber, Florian and Jia, Ningyuan and Kedar, Dhruv and Kornjača, Milan and Liu, Fangli and Long, John and Lopatin, Jonathan and Lopes, Pedro L. S. and Luo, Xiu-Zhe and Macrì, Tommaso and Marković, Ognjen and Martínez-Martínez, Luis A. and Meng, Xianmei and Ostermann, Stefan and Ostroumov, Evgeny and Paquette, David and Qiang, Zexuan and Shofman, Vadim and Singh, Anshuman and Singh, Manuj and Sinha, Nandan and Thoreen, Henry and Wan, Noel and Wang, Yiping and Waxman-Lenz, Daniel and Wong, Tak and Wurtz, Jonathan and Zhdanov, Andrii and Zheng, Laurent and Greiner, Markus and Keesling, Alexander and Gemelke, Nathan and Vuletić, Vladan and Kitagawa, Takuya and Wang, Sheng-Tao and Bluvstein, Dolev and Lukin, Mikhail D. and Lukin, Alexander and Zhou, Hengyun and Cantú, Sergio H.},
    author={Sales Rodriguez, Pedro and Robinson, John M and Jepsen, Paul Niklas and He, Zhiyang and Duckering, Casey and Zhao, Chen and Wu, Kai-Hsin and Campo, Joseph and Bagnall, Kevin and Kwon, Minho and others},
	date = {2025/09/01},
	doi = {10.1038/s41586-025-09367-3},
	id = {Sales Rodriguez2025},
	isbn = {1476-4687},
	journal = {Nature},
	number = {8081},
	pages = {620--625},
	title = {Experimental demonstration of logical magic state distillation},
	url = {https://doi.org/10.1038/s41586-025-09367-3},
	volume = {645},
	year = {2025}}

@article{MaAlex2022YbUniversalGate,
  title = {Universal Gate Operations on Nuclear Spin Qubits in an Optical Tweezer Array of $^{171}\mathrm{Yb}$ Atoms},
  author = {Ma, Shuo and Burgers, Alex P. and Liu, Genyue and Wilson, Jack and Zhang, Bichen and Thompson, Jeff D.},
  journal = {Phys. Rev. X},
  volume = {12},
  issue = {2},
  pages = {021028},
  numpages = {12},
  year = {2022},
  month = {May},
  publisher = {American Physical Society},
  doi = {10.1103/PhysRevX.12.021028},
  url = {https://link.aps.org/doi/10.1103/PhysRevX.12.021028}
}

@article{Evered2023HighFidelityEntanglingGate,
	author = {Evered, Simon J. and Bluvstein, Dolev and Kalinowski, Marcin and Ebadi, Sepehr and Manovitz, Tom and Zhou, Hengyun and Li, Sophie H. and Geim, Alexandra A. and Wang, Tout T. and Maskara, Nishad and Levine, Harry and Semeghini, Giulia and Greiner, Markus and Vuletić, Vladan and Lukin, Mikhail D.},
	journal = {Nature},
	number = {7982},
	pages = {268--272},
	title = {High-fidelity parallel entangling gates on a neutral-atom quantum computer},
	volume = {622},
	year = {2023},
    url = {https://doi.org/10.1038/s41586-023-06481-y}
}

@article{Tsai2025CZLinearResponse,
  title = {Benchmarking and Fidelity Response Theory of High-Fidelity Rydberg Entangling Gates},
  author = {Tsai, Richard Bing-Shiun and Sun, Xiangkai and Shaw, Adam L. and Finkelstein, Ran and Endres, Manuel},
  journal = {PRX Quantum},
  volume = {6},
  issue = {1},
  pages = {010331},
  numpages = {28},
  year = {2025},
  month = {Feb},
  publisher = {American Physical Society},
  doi = {10.1103/PRXQuantum.6.010331},
  url = {https://link.aps.org/doi/10.1103/PRXQuantum.6.010331}
}

@article{Kitaev1997QuantumErrorCorrection,
	author = {A Yu Kitaev},
	journal = {Russ. Math. Surv.},
	number = {6},
	pages = {1191},
	title = {Quantum computations: algorithms and error correction},
	volume = {52},
	year = {1997},
    url ={https://iopscience.iop.org/article/10.1070/RM1997v052n06ABEH002155/}
}

@article{Manuel2016Rearranging,
	author = {Manuel Endres and Hannes Bernien and Alexander Keesling and Harry Levine and Eric R. Anschuetz and Alexandre Krajenbrink and Crystal Senko and Vladan Vuletic and Markus Greiner and Mikhail D. Lukin},
	journal = {Science},
	number = {6315},
	pages = {1024-1027},
	title = {Atom-by-atom assembly of defect-free one-dimensional cold atom arrays},
	volume = {354},
	year = {2016},
    url = {https://www.science.org/doi/10.1126/science.aah3752}
}

@article{Porsev1999DataYb,
  title = {Electric-dipole amplitudes, lifetimes, and polarizabilities of the low-lying levels of atomic ytterbium},
  author = {Porsev, S. G. and Rakhlina, Yu. G. and Kozlov, M. G.},
  journal = {Phys. Rev. A},
  volume = {60},
  issue = {4},
  pages = {2781--2785},
  numpages = {0},
  year = {1999},
  month = {Oct},
  publisher = {American Physical Society},
  doi = {10.1103/PhysRevA.60.2781},
  url = {https://link.aps.org/doi/10.1103/PhysRevA.60.2781}
}

@article{li2026quantumsciencearraysmetastable,
      title={Quantum science with arrays of metastable helium-3 atoms}, 
      author={Zheyuan Li and Rupsa De and Rishi Sivakumar and William Huie and Hao-Tian Wei and Justin D. Piel and Chris H. Greene and Kaden R. A. Hazzard and Zoe Z. Yan and Jacob P. Covey},
      year={2026},
      journal={arXiv:2601.06763},
      url={https://arxiv.org/abs/2601.06763}, 
}

@article{Daniel2016Rearranging,
	author = {Daniel Barredo and Sylvain de L{\'e}s{\'e}leuc and Vincent Lienhard and Thierry Lahaye and Antoine Browaeys},
	journal = {Science},
	number = {6315},
	pages = {1021-1023},
	title = {An atom-by-atom assembler of defect-free arbitrary two-dimensional atomic arrays},
	volume = {354},
	year = {2016},
    url = {https://www.science.org/doi/10.1126/science.aah3778}
}

@article{Levine2019LevinePichlerGate,
  title = {Parallel Implementation of High-Fidelity Multiqubit Gates with Neutral Atoms},
  author = {Levine, Harry and Keesling, Alexander and Semeghini, Giulia and Omran, Ahmed and Wang, Tout T. and Ebadi, Sepehr and Bernien, Hannes and Greiner, Markus and Vuleti\ifmmode \acute{c}\else \'{c}\fi{}, Vladan and Pichler, Hannes and Lukin, Mikhail D.},
  journal = {Phys. Rev. Lett.},
  volume = {123},
  issue = {17},
  pages = {170503},
  numpages = {6},
  year = {2019},
  month = {Oct},
  publisher = {American Physical Society},
  doi = {10.1103/PhysRevLett.123.170503},
  url = {https://link.aps.org/doi/10.1103/PhysRevLett.123.170503}
}

@article{Bluvstein2022EntanglingTransport,
	author = {Bluvstein, Dolev and Levine, Harry and Semeghini, Giulia and Wang, Tout T. and Ebadi, Sepehr and Kalinowski, Marcin and Keesling, Alexander and Maskara, Nishad and Pichler, Hannes and Greiner, Markus and Vuletic, Vladan and Lukin, Mikhail D.},
	journal = {Nature},
	number = {7906},
	pages = {451--456},
	title = {A quantum processor based on coherent transport of entangled atom arrays},
	volume = {604},
	year = {2022},
    url = {https://doi.org/10.1038/s41586-022-04592-6}
}

@article{kuzmich2005atomphotonentanglement,
  title={Entanglement of a photon and a collective atomic excitation},
  author={Matsukevich, DN and Chaneliere, Thierry and Bhattacharya, Mishkatul and Lan, S-Y and Jenkins, SD and Kennedy, TAB and Kuzmich, Alex},
  journal={Phys. Rev. Lett.},
  volume={95},
  number={4},
  pages={040405},
  year={2005},
  publisher={APS},
  url = {https://doi.org/10.1103/PhysRevLett.95.040405}
}

@article{Luan2020PhotonicCrystalTrap,
	author = {Luan, Xingsheng and B{\'e}guin, Jean-Baptiste and Burgers, Alex P. and Qin, Zhongzhong and Yu, Su-Peng and Kimble, Harry J.},
	journal = {Adv. Quantum Technol.},
	number = {11},
	pages = {2000008},
	title = {The Integration of Photonic Crystal Waveguides with Atom Arrays in Optical Tweezers},
	volume = {3},
	year = {2020},
    url = {https://doi.org/10.1002/qute.202000008}
}

@article{Menon2024PhotonicCrystalTrap,
	author = {Menon, Shankar G. and Glachman, Noah and Pompili, Matteo and Dibos, Alan and Bernien, Hannes},
	journal = {Nat. Commun.},
	number = {1},
	pages = {6156},
	title = {An integrated atom array-nanophotonic chip platform with background-free imaging},
	volume = {15},
	year = {2024},
    url = {https://doi.org/10.1038/s41467-024-50355-4}
}

@article{yiyi2024ModularQPU,
  title = {High-Rate and High-Fidelity Modular Interconnects between Neutral Atom Quantum Processors},
  author = {Li, Yiyi and Thompson, Jeff D.},
  journal = {PRX Quantum},
  volume = {5},
  issue = {2},
  pages = {020363},
  numpages = {13},
  year = {2024},
  month = {Jun},
  publisher = {American Physical Society},
  doi = {10.1103/PRXQuantum.5.020363},
  url = {https://link.aps.org/doi/10.1103/PRXQuantum.5.020363}
}

@article{Li2025InterconnectNode,
	author = {Li, Lintao and Hu, Xiye and Jia, Zhubing and Huie, William and Sun, Won Kyu Calvin and Aakash and Dong, Yuhao and Hiri-O-Tuppa, Narisak and Covey, Jacob P.},
	journal = {Nat. Phys.},
	number = {11},
	pages = {1826--1833},
	title = {Parallelized telecom quantum networking with an ytterbium-171 atom array},
	volume = {21},
	year = {2025},
    url = {https://doi.org/10.1038/s41567-025-03022-4}
}

@article{Sinclair2025OpticalInterconnect,
  title = {Fault-tolerant optical interconnects for neutral-atom arrays},
  author = {Sinclair, Josiah and Ramette, Joshua and Grinkemeyer, Brandon and Bluvstein, Dolev and Lukin, Mikhail D. and Vuleti\ifmmode \acute{c}\else \'{c}\fi{}, Vladan},
  journal = {Phys. Rev. Res.},
  volume = {7},
  issue = {1},
  pages = {013313},
  numpages = {9},
  year = {2025},
  month = {Mar},
  publisher = {American Physical Society},
  doi = {10.1103/PhysRevResearch.7.013313},
  url = {https://link.aps.org/doi/10.1103/PhysRevResearch.7.013313}
}

@article{Sunami2025FiberInterconnect,
  title = {Scalable Networking of Neutral-Atom Qubits: Nanofiber-Based Approach for Multiprocessor Fault-Tolerant Quantum Computers},
  author = {Sunami, Shinichi and Tamiya, Shiro and Inoue, Ryotaro and Yamasaki, Hayata and Goban, Akihisa},
  journal = {PRX Quantum},
  volume = {6},
  issue = {1},
  pages = {010101},
  numpages = {22},
  year = {2025},
  month = {Feb},
  publisher = {American Physical Society},
  doi = {10.1103/PRXQuantum.6.010101},
  url = {https://link.aps.org/doi/10.1103/PRXQuantum.6.010101}
}

@article{Bluvstein2024QPU,
    FULLauthor = {Bluvstein, Dolev and Evered, Simon J. and Geim, Alexandra A. and Li,
    Sophie H. and Zhou, Hengyun and Manovitz, Tom and Ebadi, Sepehr and
    Cain, Madelyn and Kalinowski, Marcin and Hangleiter, Dominik and
    Ataides, J. Pablo Bonilla and Maskara, Nishad and Cong, Iris and Gao,
    Xun and Sales Rodriguez, Pedro and Karolyshyn, Thomas and Semeghini,
    Giulia and Gullans, Michael J. and Greiner, Markus and Vuletic, Vladan
    and Lukin, Mikhail D.},
	author = {Bluvstein, Dolev and Evered, Simon J and Geim, Alexandra A and Li, Sophie H and Zhou, Hengyun and Manovitz, Tom and Ebadi, Sepehr and Cain, Madelyn and Kalinowski, Marcin and Hangleiter, Dominik and others},
	journal = {Nature},
	number = {7997},
	pages = {58--65},
	title = {Logical quantum processor based on reconfigurable atom arrays},
	volume = {626},
	year = {2024},
    url = {https://doi.org/10.1038/s41586-023-06927-3}
}

@article{Bluvstein2025UiversalQPU,
  title={A fault-tolerant neutral-atom architecture for universal quantum computation},
  FULLauthor = {Bluvstein, Dolev and Geim, Alexandra A. and Li, Sophie H. and Evered, Simon J. and Bonilla Ataides, J. Pablo and Baranes, Gefen and Gu, Andi and Manovitz, Tom and Xu, Muqing and Kalinowski, Marcin and Majidy, Shayan and Kokail, Christian and Maskara, Nishad and Trapp, Elias C. and Stewart, Luke M. and Hollerith, Simon and Zhou, Hengyun and Gullans, Michael J. and Yelin, Susanne F. and Greiner, Markus and Vuletić, Vladan and Cain, Madelyn and Lukin, Mikhail D.},
  author={Bluvstein, Dolev and Geim, Alexandra A and Li, Sophie H and Evered, Simon J and Bonilla Ataides, J Pablo and Baranes, Gefen and Gu, Andi and Manovitz, Tom and Xu, Muqing and Kalinowski, Marcin and others},
  journal={Nature},
  volume={649},
  number={8095},
  pages={39--46},
  year={2026},
  publisher={Nature Publishing Group UK London},
  url = {https://doi.org/10.1038/s41586-025-09848-5}
}

@article{jones2007fasthyperfinequbit,
  title={Fast quantum state control of a single trapped neutral atom},
  author={Jones, Matthew PA and Beugnon, J{\'e}r{\^o}me and Ga{\"e}tan, Alpha and Zhang, Junxiang and Messin, Gaetan and Browaeys, Antoine and Grangier, Philippe},
  journal={Phys. Rev. A},
  volume={75},
  number={4},
  pages={040301},
  year={2007},
  publisher={APS},
  url = {https://doi.org/10.1103/PhysRevA.75.040301}
}

@article{saffman2010quantuminformation,
  title={Quantum information with Rydberg atoms},
  author={Saffman, Mark and Walker, Thad G and M{\o}lmer, Klaus},
  journal={Rev. Mod. Phys.},
  volume={82},
  number={3},
  pages={2313--2363},
  year={2010},
  publisher={APS},
  url = {https://doi.org/10.1103/RevModPhys.82.2313}
}

@article{wang2014quantumCesiumHyperfineQubit,
  title={Quantum state manipulation of single-Cesium-atom qubit in a micro-optical trap},
  author={Wang, Zhi-Hui and Li, Gang and Tian, Ya-Li and Zhang, Tian-Cai},
  journal={Front. Phys.},
  volume={9},
  number={5},
  pages={634--639},
  year={2014},
  publisher={Springer},
  url = {https://doi.org/10.1007/s11467-014-0442-0}
}

@article{noguchi2011quantum1S0coherenceNuclearSpin,
  title={Quantum-state tomography of a single nuclear spin qubit of an optically manipulated ytterbium atom},
  author={Noguchi, Atsushi and Eto, Yujiro and Ueda, Masahito and Kozuma, Mikio},
  journal={Phys. Rev. A},
  volume={84},
  number={3},
  pages={030301},
  year={2011},
  publisher={APS},
  url = {https://doi.org/10.1103/PhysRevA.84.030301}
}

@article{ludlow2015opticallatticeNuclearSpin,
  title={Optical atomic clocks},
  author={Ludlow, Andrew D and Boyd, Martin M and Ye, Jun and Peik, Ekkehard and Schmidt, Piet O},
  journal={Rev. Mod. Phys.},
  volume={87},
  number={2},
  pages={637--701},
  year={2015},
  publisher={APS},
  url = {https://doi.org/10.1103/RevModPhys.87.637}
}

@article{Cooper2018SrArray,
  title = {Alkaline-Earth Atoms in Optical Tweezers},
  author = {Cooper, Alexandre and Covey, Jacob P. and Madjarov, Ivaylo S. and Porsev, Sergey G. and Safronova, Marianna S. and Endres, Manuel},
  journal = {Phys. Rev. X},
  volume = {8},
  issue = {4},
  pages = {041055},
  numpages = {19},
  year = {2018},
  month = {Dec},
  publisher = {American Physical Society},
  doi = {10.1103/PhysRevX.8.041055},
  url = {https://link.aps.org/doi/10.1103/PhysRevX.8.041055}
}

@article{Saskin2019Ybqubit,
  title = {Narrow-Line Cooling and Imaging of Ytterbium Atoms in an Optical Tweezer Array},
  author = {Saskin, S. and Wilson, J. T. and Grinkemeyer, B. and Thompson, J. D.},
  journal = {Phys. Rev. Lett.},
  volume = {122},
  issue = {14},
  pages = {143002},
  numpages = {6},
  year = {2019},
  month = {Apr},
  publisher = {American Physical Society},
  doi = {10.1103/PhysRevLett.122.143002},
  url = {https://link.aps.org/doi/10.1103/PhysRevLett.122.143002}
}

@article{Jenkins2022Ybqubit,
  title = {Ytterbium Nuclear-Spin Qubits in an Optical Tweezer Array},
  author = {Jenkins, Alec and Lis, Joanna W. and Senoo, Aruku and McGrew, William F. and Kaufman, Adam M.},
  journal = {Phys. Rev. X},
  volume = {12},
  issue = {2},
  pages = {021027},
  numpages = {21},
  year = {2022},
  month = {May},
  publisher = {American Physical Society},
  doi = {10.1103/PhysRevX.12.021027},
  url = {https://link.aps.org/doi/10.1103/PhysRevX.12.021027}
}

@article{Wu2022ErasureConversionAEA,
	author = {Wu, Yue and Kolkowitz, Shimon and Puri, Shruti and Thompson, Jeff D.},
	journal = {Nat. Commun.},
	number = {1},
	pages = {4657},
	title = {Erasure conversion for fault-tolerant quantum computing in alkaline earth Rydberg atom arrays},
	volume = {13},
	year = {2022},
    url = {https://doi.org/10.1038/s41467-022-32094-6}
}

@article{Singh2023Midcircuit,
  title={Mid-circuit correction of correlated phase errors using an array of spectator qubits},
  author={Singh, Kevin and Bradley, Conor E and Anand, Shraddha and Ramesh, Vikram and White, Ryan and Bernien, Hannes},
  journal={Science},
  volume={380},
  number={6651},
  pages={1265--1269},
  year={2023},
  publisher={American Association for the Advancement of Science},
  url={https://www.science.org/doi/10.1126/science.ade5337}
}

@article{Graham2023Midcircuit,
  title = {Midcircuit Measurements on a Single-Species Neutral Alkali Atom Quantum Processor},
  author = {Graham, T. M. and Phuttitarn, L. and Chinnarasu, R. and Song, Y. and Poole, C. and Jooya, K. and Scott, J. and Scott, A. and Eichler, P. and Saffman, M.},
  journal = {Phys. Rev. X},
  volume = {13},
  issue = {4},
  pages = {041051},
  numpages = {22},
  year = {2023},
  month = {Dec},
  publisher = {American Physical Society},
  doi = {10.1103/PhysRevX.13.041051},
  url = {https://link.aps.org/doi/10.1103/PhysRevX.13.041051}
}

@article{Norcia2023Midcircuit,
  title = {Midcircuit Qubit Measurement and Rearrangement in a $^{171}\mathrm{Yb}$ Atomic Array},
  FULLauthor = {Norcia, M. A. and Cairncross, W. B. and Barnes, K. and Battaglino, P. and Brown, A. and Brown, M. O. and Cassella, K. and Chen, C.-A. and Coxe, R. and Crow, D. and Epstein, J. and Griger, C. and Jones, A. M. W. and Kim, H. and Kindem, J. M. and King, J. and Kondov, S. S. and Kotru, K. and Lauigan, J. and Li, M. and Lu, M. and Megidish, E. and Marjanovic, J. and McDonald, M. and Mittiga, T. and Muniz, J. A. and Narayanaswami, S. and Nishiguchi, C. and Notermans, R. and Paule, T. and Pawlak, K. A. and Peng, L. S. and Ryou, A. and Smull, A. and Stack, D. and Stone, M. and Sucich, A. and Urbanek, M. and van de Veerdonk, R. J. M. and Vendeiro, Z. and Wilkason, T. and Wu, T.-Y. and Xie, X. and Zhang, X. and Bloom, B. J.},
  author = {Norcia, MA and Cairncross, WB and Barnes, K and Battaglino, P and Brown, A and Brown, MO and Cassella, K and Chen, C-A and Coxe, R and Crow, D and others},
  journal = {Phys. Rev. X},
  volume = {13},
  issue = {4},
  pages = {041034},
  numpages = {12},
  year = {2023},
  month = {Nov},
  publisher = {American Physical Society},
  doi = {10.1103/PhysRevX.13.041034},
  url = {https://link.aps.org/doi/10.1103/PhysRevX.13.041034}
}

@article{Ma2023YbMidcircuitErasure,
  title={High-fidelity gates and mid-circuit erasure conversion in an atomic qubit},
  author={Ma, Shuo and Liu, Genyue and Peng, Pai and Zhang, Bichen and Jandura, Sven and Claes, Jahan and Burgers, Alex P and Pupillo, Guido and Puri, Shruti and Thompson, Jeff D},
  journal={Nature},
  volume={622},
  number={7982},
  pages={279--284},
  year={2023},
  publisher={Nature Publishing Group UK London},
  url = {https://doi.org/10.1038/s41586-023-06438-1}
}

@article{zhang2025leveragingerasureerrorslogical,
      title={Leveraging erasure errors in logical qubits with metastable $^{171}\mathrm{Yb}$ atoms},
      author={Zhang, Bichen and Liu, Genyue and Bornet, Guillaume and Horvath, Sebastian P and Peng, Pai and Ma, Shuo and Huang, Shilin and Puri, Shruti and Thompson, Jeff D},
      journal={arXiv:2506.13724},
      year={2025},
      url={https://arxiv.org/abs/2506.13724}, 
}

@article{evered2026highfidelityentanglinggatesnonlocal,
  title={High-fidelity entangling gates and nonlocal circuits with neutral atoms},
  author={Evered, Simon J and Xu, Muqing and Li, Sophie H and Geim, Alexandra A and Ataides, J and Kalinowski, Marcin and Bluvstein, Dolev and Maskara, Nishad and Kokail, Christian and Greiner, Markus and others},
  journal={arXiv:2604.25987},
  year={2026},
  url={https://arxiv.org/abs/2604.25987}, 
}

@article{Radnaev2025UnivQPULocalDressing,
  title = {Universal Neutral-Atom Quantum Computer with Individual Optical Addressing and Nondestructive Readout},
  FULLauthor = {Radnaev, A.G. and Chung, W.C. and Cole, D.C. and Mason, D. and Ballance, T.G. and Bedalov, M.J. and Belknap, D.A. and Berman, M.R. and Blakely, M. and Bloomfield, I.L. and Buttler, P.D. and Campbell, C. and Chopinaud, A. and Copenhaver, E. and Dawes, M.K. and Eubanks, S.Y. and Friss, A.J. and Garcia, D.M. and Gilbert, J. and Gillette, M. and Goiporia, P. and Gokhale, P. and Goldwin, J. and Goodwin, D. and Graham, T.M. and Guttormsson, C.J. and Hickman, G.T. and Hurtley, L. and Iliev, M. and Jones, E.B. and Jones, R.A. and Kuper, K.W. and Lewis, T.B. and Lichtman, M.T. and Majdeteimouri, F. and Mason, J.J. and McMaster, J.K. and Miles, J.A. and Mitchell, P.T. and Murphree, J.D. and Neff-Mallon, N.A. and Oh, T. and Omole, V. and Parlo Simon, C. and Pederson, N. and Perlin, M.A. and Reiter, A. and Rines, R. and Romlow, P. and Scott, A.M. and Stiefvater, D. and Tanner, J.R. and Tucker, A.K. and Vinogradov, I.V. and Warter, M.L. and Yeo, M. and Saffman, M. and Noel, T.W.},
  author={Radnaev, AG and Chung, WC and Cole, DC and Mason, D and Ballance, TG and Bedalov, MJ and Belknap, DA and Berman, MR and Blakely, M and Bloomfield, IL and others},
  journal = {PRX Quantum},
  volume = {6},
  issue = {3},
  pages = {030334},
  numpages = {20},
  year = {2025},
  month = {Aug},
  publisher = {American Physical Society},
  doi = {10.1103/66s8-jj18},
  url = {https://link.aps.org/doi/10.1103/66s8-jj18}
}

@article{Preskill2018quantumcomputingin,
  doi = {10.22331/q-2018-08-06-79},
  url = {https://doi.org/10.22331/q-2018-08-06-79},
  title = {Quantum {C}omputing in the {NISQ} era and beyond},
  author = {Preskill, John},
  journal = {{Quantum}},
  issn = {2521-327X},
  publisher = {{Verein zur F{\"{o}}rderung des Open Access Publizierens in den Quantenwissenschaften}},
  volume = {2},
  pages = {79},
  month = aug,
  year = {2018}
}

@article{Preskill2025MegaGroup,
    title={Beyond nisq: The megaquop machine},
    author={Preskill, John},
    journal={ACM Trans. Quant. Comp.},
    volume={6},
    number={3},
    pages={1--7},
    year={2025},
    publisher={ACM New York},
    url = {https://doi.org/10.1145/3723153},
}

@article{draper2004LogDepthOptimzation,
  title={A logarithmic-depth quantum carry-lookahead adder},
  author={Draper, Thomas G and Kutin, Samuel A and Rains, Eric M and Svore, Krysta M},
  journal={arXiv: quant-ph/0406142},
  year={2004},
  url = {https://doi.org/10.48550/arXiv.quant-ph/0406142}
}

@article{Sahay2023ErasureMakeHighTresholdCodes,
  title = {High-Threshold Codes for Neutral-Atom Qubits with Biased Erasure Errors},
  author = {Sahay, Kaavya and Jin, Junlan and Claes, Jahan and Thompson, Jeff D. and Puri, Shruti},
  journal = {Phys. Rev. X},
  volume = {13},
  issue = {4},
  pages = {041013},
  numpages = {12},
  year = {2023},
  month = {Oct},
  publisher = {American Physical Society},
  doi = {10.1103/PhysRevX.13.041013},
  url = {https://link.aps.org/doi/10.1103/PhysRevX.13.041013}
}

@article{Steane1996SteaneCode,
  title = {Error Correcting Codes in Quantum Theory},
  author = {Steane, A. M.},
  journal = {Phys. Rev. Lett.},
  volume = {77},
  issue = {5},
  pages = {793--797},
  numpages = {0},
  year = {1996},
  month = {Jul},
  publisher = {American Physical Society},
  doi = {10.1103/PhysRevLett.77.793},
  url = {https://link.aps.org/doi/10.1103/PhysRevLett.77.793}
}

@book{gottesman1997stabilizercodesquantumerror,
        title={Stabilizer codes and quantum error correction},
        author={Gottesman, Daniel},
        year={1997},
        publisher={California Institute of Technology}
}

@article{Fowler2012SurfaceCode,
  title = {Surface codes: Towards practical large-scale quantum computation},
  author = {Fowler, Austin G. and Mariantoni, Matteo and Martinis, John M. and Cleland, Andrew N.},
  journal = {Phys. Rev. A},
  volume = {86},
  issue = {3},
  pages = {032324},
  numpages = {48},
  year = {2012},
  month = {Sep},
  publisher = {American Physical Society},
  doi = {10.1103/PhysRevA.86.032324},
  url = {https://link.aps.org/doi/10.1103/PhysRevA.86.032324}
}

@article{Kjaergaard2020SCqubitConcept,
   title={Superconducting Qubits: Current State of Play},
   volume={11},
   ISSN={1947-5462},
   url={http://dx.doi.org/10.1146/annurev-conmatphys-031119-050605},
   DOI={10.1146/annurev-conmatphys-031119-050605},
   number={1},
   journal={Annu. Rev. Condens. Matter Phys.},
   publisher={Annual Reviews},
   author={Kjaergaard, Morten and Schwartz, Mollie E. and Braumüller, Jochen and Krantz, Philip and Wang, Joel I.-J. and Gustavsson, Simon and Oliver, William D.},
   year={2020},
   month=mar, pages={369–395} 
}

@article{Deist2022CavityMidcircuit,
  title = {Mid-Circuit Cavity Measurement in a Neutral Atom Array},
  author = {Deist, Emma and Lu, Yue-Hui and Ho, Jacquelyn and Pasha, Mary Kate and Zeiher, Johannes and Yan, Zhenjie and Stamper-Kurn, Dan M.},
  journal = {Phys. Rev. Lett.},
  volume = {129},
  issue = {20},
  pages = {203602},
  numpages = {7},
  year = {2022},
  month = {Nov},
  publisher = {American Physical Society},
  doi = {10.1103/PhysRevLett.129.203602},
  url = {https://link.aps.org/doi/10.1103/PhysRevLett.129.203602}
}

@article{anand2024dualspiciesBernien,
  title={A dual-species Rydberg array},
  author={Anand, Shraddha and Bradley, Conor E and White, Ryan and Ramesh, Vikram and Singh, Kevin and Bernien, Hannes},
  journal={Nat. Phys.},
  volume={20},
  number={11},
  pages={1744--1750},
  year={2024},
  publisher={Nature Publishing Group UK London},
  url = {https://doi.org/10.1038/s41567-024-02638-2}
}

@article{petrosyan2024fastDualSpeciesSaffman,
  title={Fast measurements and multiqubit gates in dual-species atomic arrays},
  author={Petrosyan, D and Norrell, S and Poole, C and Saffman, M},
  journal={Phys. Rev. A},
  volume={110},
  number={4},
  pages={042404},
  year={2024},
  publisher={APS},
  url = {https://doi.org/10.1103/PhysRevA.110.042404}
}

@article{zhang2025dualspeciesAQDQ,
  title={Dual-type dual-element atom arrays for quantum information processing},
  author={Zhang, Zhanchuan and Arunseangroj, Jeth and Xu, Wenchao},
  journal={arXiv:2503.16896},
  year={2025},
  url = {https://doi.org/10.48550/arXiv.2503.16896}
}

@article{white2026quantumcellularautomatadualspecies,
  title={Quantum Cellular Automata on a Dual-Species Rydberg Processor},
  author={White, Ryan and Ramesh, Vikram and Impertro, Alexander and Anand, Shraddha and Cesa, Francesco and Giudici, Giuliano and Iadecola, Thomas and Pichler, Hannes and Bernien, Hannes},
  journal={arXiv:2601.16257},
  year={2026},
  url={https://arxiv.org/abs/2601.16257}, 
}

@article{zeng2017entanglingDualIsotope,
  title={Entangling two individual atoms of different isotopes via Rydberg blockade},
  author={Zeng, Yong and Xu, Peng and He, Xiaodong and Liu, Yangyang and Liu, Min and Wang, Jin and Papoular, DJ and Shlyapnikov, GV and Zhan, Mingsheng},
  journal={Phys. Rev. Lett.},
  volume={119},
  number={16},
  pages={160502},
  year={2017},
  publisher={APS},
  url = {https://doi.org/10.1103/PhysRevLett.119.160502}
}

@article{nakamura2024hybridDualisotopeAQDQ,
  title={Hybrid atom tweezer array of nuclear spin and optical clock qubits},
  author={Nakamura, Yuma and Kusano, Toshi and Yokoyama, Rei and Saito, Keito and Higashi, Koichiro and Ozawa, Naoya and Takano, Tetsushi and Takasu, Yosuke and Takahashi, Yoshiro},
  journal={Phys. Rev. X},
  volume={14},
  number={4},
  pages={041062},
  year={2024},
  publisher={APS},
  url = {https://doi.org/10.1103/PhysRevX.14.041062}
}

@article{yang2022dualtypeion,
  title={Realizing coherently convertible dual-type qubits with the same ion species},
  author={Yang, H-X and Ma, J-Y and Wu, Y-K and Wang, Ye and Cao, M-M and Guo, W-X and Huang, Y-Y and Feng, Lu and Zhou, Z-C and Duan, L-M},
  journal={Nat. Phys.},
  volume={18},
  number={9},
  pages={1058--1061},
  year={2022},
  publisher={Nature Publishing Group UK London},
  url = {https://doi.org/10.1038/s41567-022-01661-5}
}

@article{Lis2023MidcircuitOMG,
      title={Midcircuit operations using the omg architecture in neutral atom arrays},
      author={Lis, Joanna W and Senoo, Aruku and McGrew, William F and R{\"o}nchen, Felix and Jenkins, Alec and Kaufman, Adam M},
      journal={Phys. Rev. X},
      volume={13},
      number={4},
      pages={041035},
      year={2023},
      publisher={APS},
      url = {https://doi.org/10.1103/PhysRevX.13.041035}
}

@article{tao2025universalgatesmetastablequbit,
      title={Universal gates for a metastable qubit in strontium-88},
      author={Tao, Renhao and Lib, Ohad and Gyger, Flavien and Timme, Hendrik and Ammenwerth, Maximilian and Bloch, Immanuel and Zeiher, Johannes},
      journal={arXiv:2506.10714},
      year={2025},
      url={https://arxiv.org/abs/2506.10714}
}

@article{Scholl2023ErasureRydbergSimulator,
	author = {Scholl, Pascal and Shaw, Adam L. and Tsai, Richard Bing-Shiun and Finkelstein, Ran and Choi, Joonhee and Endres, Manuel},
	journal = {Nature},
	number = {7982},
	pages = {273--278},
	title = {Erasure conversion in a high-fidelity Rydberg quantum simulator},
	volume = {622},
	year = {2023},
    url = {https://doi.org/10.1038/s41586-023-06516-4}
}

@article{Adam2025ErasureCoolingMotionalState,
	author = {Adam L. Shaw and Pascal Scholl and Ran Finkelstein and Richard Bing-Shiun Tsai and Joonhee Choi and Manuel Endres},
	journal = {Science},
	number = {6749},
	pages = {845-849},
	title = {Erasure cooling, control, and hyperentanglement of motion in optical tweezers},
	volume = {388},
	year = {2025},
    url = {https://www.science.org/doi/10.1126/science.adn2618}
}

@article{Knill2008RandomizedBenchmark,
  title = {Randomized benchmarking of quantum gates},
  author = {Knill, E. and Leibfried, D. and Reichle, R. and Britton, J. and Blakestad, R. B. and Jost, J. D. and Langer, C. and Ozeri, R. and Seidelin, S. and Wineland, D. J.},
  journal = {Phys. Rev. A},
  volume = {77},
  issue = {1},
  pages = {012307},
  numpages = {7},
  year = {2008},
  month = {Jan},
  publisher = {American Physical Society},
  doi = {10.1103/PhysRevA.77.012307},
  url = {https://link.aps.org/doi/10.1103/PhysRevA.77.012307}
}

@article{Wineland:92,
author = {D. J. Wineland and J. Dalibard and C. Cohen-Tannoudji},
journal = {J. Opt. Soc. Am. B},
number = {1},
pages = {32--42},
publisher = {Optica Publishing Group},
title = {Sisyphus cooling of a bound atom},
volume = {9},
month = {Jan},
year = {1992},
url = {https://opg.optica.org/josab/abstract.cfm?URI=josab-9-1-32},
doi = {10.1364/JOSAB.9.000032},
}

@article{PhysRevLett.131.083001,
  title = {Imaging a $^{6}\mathrm{Li}$ Atom in an Optical Tweezer 2000 Times with $\mathrm{\ensuremath{\Lambda}}$-Enhanced Gray Molasses},
  author = {Blodgett, Karl N. and Peana, David and Phatak, Saumitra S. and Terry, Lane M. and Montes, Maria Paula and Hood, Jonathan D.},
  journal = {Phys. Rev. Lett.},
  volume = {131},
  issue = {8},
  pages = {083001},
  numpages = {5},
  year = {2023},
  month = {Aug},
  publisher = {American Physical Society},
  doi = {10.1103/PhysRevLett.131.083001},
  url = {https://link.aps.org/doi/10.1103/PhysRevLett.131.083001}
}

@article{Kuhr2005DephasingTime,
  title = {Analysis of dephasing mechanisms in a standing-wave dipole trap},
  author = {Kuhr, S. and Alt, W. and Schrader, D. and Dotsenko, I. and Miroshnychenko, Y. and Rauschenbeutel, A. and Meschede, D.},
  journal = {Phys. Rev. A},
  volume = {72},
  issue = {2},
  pages = {023406},
  numpages = {12},
  year = {2005},
  month = {Aug},
  publisher = {American Physical Society},
  doi = {10.1103/PhysRevA.72.023406},
  url = {https://link.aps.org/doi/10.1103/PhysRevA.72.023406}
}

@article{Beugnon2007TransferQubit,
  title={Two-dimensional transport and transfer of a single atomic qubit in optical tweezers},
  author={Beugnon, J{\'e}r{\^o}me and Tuchendler, Charles and Marion, Harold and Ga{\"e}tan, Alpha and Miroshnychenko, Yevhen and Sortais, Yvan RP and Lance, Andrew M and Jones, Matthew PA and Messin, Gaetan and Browaeys, Antoine and Grangier, Philippe},
  journal={Nat. Phys.},
  volume={3},
  number={10},
  pages={696--699},
  year={2007},
  publisher={Nature Publishing Group UK London},
  url = {https://www.nature.com/articles/nphys698}
}

@article{Levine2022DispersiveHyperfineQubit,
  title = {Dispersive optical systems for scalable Raman driving of hyperfine qubits},
  author = {Levine, Harry and Bluvstein, Dolev and Keesling, Alexander and Wang, Tout T. and Ebadi, Sepehr and Semeghini, Giulia and Omran, Ahmed and Greiner, Markus and Vuleti\ifmmode \acute{c}\else \'{c}\fi{}, Vladan and Lukin, Mikhail D.},
  journal = {Phys. Rev. A},
  volume = {105},
  issue = {3},
  pages = {032618},
  numpages = {13},
  year = {2022},
  month = {Mar},
  publisher = {American Physical Society},
  doi = {10.1103/PhysRevA.105.032618},
  url = {https://link.aps.org/doi/10.1103/PhysRevA.105.032618}
}

@article{Xia2015RBSingleQubit,
  title = {Randomized Benchmarking of Single-Qubit Gates in a 2D Array of Neutral-Atom Qubits},
  author = {Xia, T. and Lichtman, M. and Maller, K. and Carr, A. W. and Piotrowicz, M. J. and Isenhower, L. and Saffman, M.},
  journal = {Phys. Rev. Lett.},
  volume = {114},
  issue = {10},
  pages = {100503},
  numpages = {5},
  year = {2015},
  month = {Mar},
  publisher = {American Physical Society},
  doi = {10.1103/PhysRevLett.114.100503},
  url = {https://link.aps.org/doi/10.1103/PhysRevLett.114.100503}
}

@article{Guo2020CoherenceTimes,
  title = {Balanced Coherence Times of Atomic Qubits of Different Species in a Dual $3\ifmmode\times\else\texttimes\fi{}3$ Magic-Intensity Optical Dipole Trap Array},
  author = {Guo, Ruijun and He, Xiaodong and Sheng, Cheng and Yang, Jiaheng and Xu, Peng and Wang, Kunpeng and Zhong, Jiaqi and Liu, Min and Wang, Jin and Zhan, Mingsheng},
  journal = {Phys. Rev. Lett.},
  volume = {124},
  issue = {15},
  pages = {153201},
  numpages = {6},
  year = {2020},
  month = {Apr},
  publisher = {American Physical Society},
  doi = {10.1103/PhysRevLett.124.153201},
  url = {https://link.aps.org/doi/10.1103/PhysRevLett.124.153201}
}

@article{Magesan2012InterleavedRB,
  title = {Efficient Measurement of Quantum Gate Error by Interleaved Randomized Benchmarking},
  author = {Magesan, Easwar and Gambetta, Jay M. and Johnson, B. R. and Ryan, Colm A. and Chow, Jerry M. and Merkel, Seth T. and da Silva, Marcus P. and Keefe, George A. and Rothwell, Mary B. and Ohki, Thomas A. and Ketchen, Mark B. and Steffen, M.},
  journal = {Phys. Rev. Lett.},
  volume = {109},
  issue = {8},
  pages = {080505},
  numpages = {5},
  year = {2012},
  month = {Aug},
  publisher = {American Physical Society},
  doi = {10.1103/PhysRevLett.109.080505},
  url = {https://link.aps.org/doi/10.1103/PhysRevLett.109.080505}
}

@article{Xu2024qLDPC,
	author = {Xu, Qian and Bonilla Ataides, J. Pablo and Pattison, Christopher A. and Raveendran, Nithin and Bluvstein, Dolev and Wurtz, Jonathan and Vasi{\'c}, Bane and Lukin, Mikhail D. and Jiang, Liang and Zhou, Hengyun},
	date = {2024/07/01},
	doi = {10.1038/s41567-024-02479-z},
	id = {Xu2024},
	isbn = {1745-2481},
	journal = {Nat. Phys.},
	number = {7},
	pages = {1084--1090},
	title = {Constant-overhead fault-tolerant quantum computation with reconfigurable atom arrays},
	url = {https://doi.org/10.1038/s41567-024-02479-z},
	volume = {20},
	year = {2024},
}

@article{Nielsen2020-rd,
  title={Probing quantum processor performance with pyGSTi},
  author={Nielsen, Erik and Rudinger, Kenneth and Proctor, Timothy and Russo, Antonio and Young, Kevin and Blume-Kohout, Robin},
  journal={Quantum Sci. Technol.},
  volume={5},
  number={4},
  pages={044002},
  year={2020},
  publisher={IOP Publishing},
  url={https://iopscience.iop.org/article/10.1088/2058-9565/ab8aa4},
}

@article{McKay2017EfficientZgate,
  title = {Efficient $Z$ gates for quantum computing},
  author = {McKay, David C. and Wood, Christopher J. and Sheldon, Sarah and Chow, Jerry M. and Gambetta, Jay M.},
  journal = {Phys. Rev. A},
  volume = {96},
  issue = {2},
  pages = {022330},
  numpages = {8},
  year = {2017},
  month = {Aug},
  publisher = {American Physical Society},
  doi = {10.1103/PhysRevA.96.022330},
  url = {https://link.aps.org/doi/10.1103/PhysRevA.96.022330}
}

@article{Graham2022MultiQubitEntangling,
	FULLauthor = {Graham, T. M. and Song, Y. and Scott, J. and Poole, C. and Phuttitarn, L. and Jooya, K. and Eichler, P. and Jiang, X. and Marra, A. and Grinkemeyer, B. and Kwon, M. and Ebert, M. and Cherek, J. and Lichtman, M. T. and Gillette, M. and Gilbert, J. and Bowman, D. and Ballance, T. and Campbell, C. and Dahl, E. D. and Crawford, O. and Blunt, N. S. and Rogers, B. and Noel, T. and Saffman, M.},
    author={Graham, TM and Song, Y and Scott, J and Poole, C and Phuttitarn, L and Jooya, K and Eichler, P and Jiang, X and Marra, A and Grinkemeyer, B and others},
	date = {2022/04/01},
	doi = {10.1038/s41586-022-04603-6},
	id = {Graham2022},
	isbn = {1476-4687},
	journal = {Nature},
	number = {7906},
	pages = {457--462},
	title = {Multi-qubit entanglement and algorithms on a neutral-atom quantum computer},
	url = {https://doi.org/10.1038/s41586-022-04603-6},
	volume = {604},
	year = {2022},
}

@article{Wang2025IndividualControlPhase,
  title = {Individual-atom control in an array through phase modulation},
  author = {Wang, Guoqing and Xu, Wenchao and Li, Changhao and Vuleti\ifmmode \acute{c}\else \'{c}\fi{}, Vladan and Cappellaro, Paola},
  journal = {Phys. Rev. Appl.},
  volume = {23},
  issue = {2},
  pages = {024072},
  numpages = {13},
  year = {2025},
  month = {Feb},
  publisher = {American Physical Society},
  doi = {10.1103/PhysRevApplied.23.024072},
  url = {https://link.aps.org/doi/10.1103/PhysRevApplied.23.024072}
}

@article{Cain2024CorrelatedDecodingTransversalGate,
  title = {Correlated Decoding of Logical Algorithms with Transversal Gates},
  author = {Cain, Madelyn and Zhao, Chen and Zhou, Hengyun and Meister, Nadine and Ataides, J. Pablo Bonilla and Jaffe, Arthur and Bluvstein, Dolev and Lukin, Mikhail D.},
  journal = {Phys. Rev. Lett.},
  volume = {133},
  issue = {24},
  pages = {240602},
  numpages = {7},
  year = {2024},
  month = {Dec},
  publisher = {American Physical Society},
  doi = {10.1103/PhysRevLett.133.240602},
  url = {https://link.aps.org/doi/10.1103/PhysRevLett.133.240602}
}

@article{Ismail2026StarArchitecture,
  title = {Transversal architecture for megaquop-scale quantum simulation with neutral atoms},
  author = {Ismail, Refaat and Chen, I-Chi and Zhao, Chen and Weiss, Ronen and Liu, Fangli and Zhou, Hengyun and Wang, Sheng-Tao and Sornborger, Andrew and Kornjača, Milan},
  journal = {PRX Quantum},
  pages = {},
  year = {2026},
  month = {Apr},
  publisher = {American Physical Society},
  doi = {10.1103/j2fw-ccmy},
  url = {https://link.aps.org/doi/10.1103/j2fw-ccmy}
}

@inproceedings{Lin2025NeutralAtomZAC,
  title={Reuse-aware compilation for zoned quantum architectures based on neutral atoms},
  author={Lin, Wan-Hsuan and Tan, Daniel Bochen and Cong, Jason},
  booktitle={2025 IEEE International Symposium on High Performance Computer Architecture (HPCA)},
  pages={127--142},
  year={2025},
  organization={IEEE}
}

@inproceedings{Stade2025RoutingAware,
  title={Routing-aware placement for zoned neutral atom-based quantum computing},
  author={Stade, Yannick and Lin, Wan-Hsuan and Cong, Jason and Wille, Robert},
  booktitle={2025 IEEE/ACM International Conference On Computer Aided Design (ICCAD)},
  pages={1--9},
  year={2025},
  organization={IEEE}
}

@article{Nakamura2025Rydberg3P2,
  title={Development of a high-power ultraviolet laser system and observation of fast coherent Rydberg excitation of ytterbium},
  author={Nakamura, Yuma and Ozawa, Naoya and Kusano, Toshi and Yokoyama, Rei and Shibata, Kosuke and Takano, Tetsushi and Takasu, Yosuke and Takahashi, Yoshiro},
  journal={J. Phys. Soc. Jpn.},
  volume={94},
  number={1},
  pages={014301},
  year={2025},
  publisher={The Physical Society of Japan},
  url={https://doi.org/10.7566/JPSJ.94.014301}
}

@article{Pepper2025MBQT,
  title = {Spectroscopy and Modeling of $^{171}\mathrm{Yb}$ Rydberg States for High-Fidelity Two-Qubit Gates},
  author = {Peper, Michael and Li, Yiyi and Knapp, Daniel Y. and Bileska, Mila and Ma, Shuo and Liu, Genyue and Peng, Pai and Zhang, Bichen and Horvath, Sebastian P. and Burgers, Alex P. and Thompson, Jeff D.},
  journal = {Phys. Rev. X},
  volume = {15},
  issue = {1},
  pages = {011009},
  numpages = {30},
  year = {2025},
  month = {Jan},
  publisher = {American Physical Society},
  doi = {10.1103/PhysRevX.15.011009},
  url = {https://link.aps.org/doi/10.1103/PhysRevX.15.011009}
}

@article{Kuroda2025PRA,
  title = {Microwave spectroscopy and multichannel quantum defect analysis of ytterbium $6snp$, $6snf$, and $6sng$ Rydberg states},
  author = {Kuroda, Rin and Hughes, Vernon M. and Poitrinal, Martin and Peper, Michael and Thompson, Jeff D.},
  journal = {Phys. Rev. A},
  volume = {112},
  issue = {4},
  pages = {042817},
  numpages = {14},
  year = {2025},
  month = {Oct},
  publisher = {American Physical Society},
  doi = {10.1103/mzsv-rckx},
  url = {https://link.aps.org/doi/10.1103/mzsv-rckx}
}

@article{Jandura2022timeoptimaltwothree,
  doi = {10.22331/q-2022-05-13-712},
  url = {https://doi.org/10.22331/q-2022-05-13-712},
  title = {Time-{O}ptimal {T}wo- and {T}hree-{Q}ubit {G}ates for {R}ydberg {A}toms},
  author = {Jandura, Sven and Pupillo, Guido},
  journal = {{Quantum}},
  issn = {2521-327X},
  publisher = {{Verein zur F{\"{o}}rderung des Open Access Publizierens in den Quantenwissenschaften}},
  volume = {6},
  pages = {712},
  month = may,
  year = {2022}
}

@article{Jandura2023OptimizingCZDesign,
  title = {Optimizing Rydberg Gates for Logical-Qubit Performance},
  author = {Jandura, Sven and Thompson, Jeff D. and Pupillo, Guido},
  journal = {PRX Quantum},
  volume = {4},
  issue = {2},
  pages = {020336},
  numpages = {17},
  year = {2023},
  month = {Jun},
  publisher = {American Physical Society},
  doi = {10.1103/PRXQuantum.4.020336},
  url = {https://link.aps.org/doi/10.1103/PRXQuantum.4.020336}
}

@article{locher2025multiqubitrydberggatesquantum,
      title={Multiqubit Rydberg Gates for Quantum Error Correction},
      author={Locher, David F and Old, Josias and Brechtelsbauer, Katharina and Holschbach, Jakob and B{\"u}chler, Hans Peter and Weber, Sebastian and M{\"u}ller, Markus},
      journal={arXiv:2512.00843},
      year={2025},
      url={https://arxiv.org/abs/2512.00843}, 
}

@article{Leseleuc2018RydbergQMCSimulation,
  title = {Analysis of imperfections in the coherent optical excitation of single atoms to Rydberg states},
  author = {de L\'es\'eleuc, Sylvain and Barredo, Daniel and Lienhard, Vincent and Browaeys, Antoine and Lahaye, Thierry},
  journal = {Phys. Rev. A},
  volume = {97},
  issue = {5},
  pages = {053803},
  numpages = {9},
  year = {2018},
  month = {May},
  publisher = {American Physical Society},
  doi = {10.1103/PhysRevA.97.053803},
  url = {https://link.aps.org/doi/10.1103/PhysRevA.97.053803}
}

@article{Pagano2022ErrorBudgeting,
  title = {Error budgeting for a controlled-phase gate with strontium-88 Rydberg atoms},
  author = {Pagano, Alice and Weber, Sebastian and Jaschke, Daniel and Pfau, Tilman and Meinert, Florian and Montangero, Simone and B\"uchler, Hans Peter},
  journal = {Phys. Rev. Res.},
  volume = {4},
  issue = {3},
  pages = {033019},
  numpages = {10},
  year = {2022},
  month = {Jul},
  publisher = {American Physical Society},
  doi = {10.1103/PhysRevResearch.4.033019},
  url = {https://link.aps.org/doi/10.1103/PhysRevResearch.4.033019}
}

@phdthesis{AdamShaw2025SrRandomness,
  author  = {Adam Lawrence Shaw},
  title   = {Learning, verifying,and erasing errors on a chaotic
and highlyentangled programmable quantum simulator},
  school  = {California Institute of Technology, Department of Physics},
  year = {2024},
  type = {Ph.D. thesis}
}

@article{senoo2025highfidelityentanglementcoherentmultiqubit,
  title={High-fidelity entanglement and coherent multi-qubit mapping in an atom array},
  author={Senoo, Aruku and Baumg{\~A}{\=I}rtner, Alexander and Lis, Joanna W and Vaidya, Gaurav M and Zeng, Zhongda and Giudici, Giuliano and Pichler, Hannes and Kaufman, Adam M},
  journal={arXiv:2506.13632},
  year={2025},
  url={https://arxiv.org/abs/2506.13632}
}

@inproceedings{Zhou2025ResourceAnalysis,
  title={Resource analysis of low-overhead transversal architectures for reconfigurable atom arrays},
  author={Zhou, Hengyun and Duckering, Casey and Zhao, Chen and Bluvstein, Dolev and Cain, Madelyn and Kubica, Aleksander and Wang, Sheng-Tao and Lukin, Mikhail D},
  booktitle={Proceedings of the 52nd Annual International Symposium on Computer Architecture},
  pages={1432--1448},
  year={2025}
}

@article{Perrin2025QECAtomLoss,
  title={Quantum error correction resilient against atom loss},
  author={Perrin, Hugo and Jandura, Sven and Pupillo, Guido},
  journal={Quantum},
  volume={9},
  pages={1884},
  year={2025},
  publisher={Verein zur F{\"o}rderung des Open Access Publizierens in den Quantenwissenschaften},
  url = {	https://doi.org/10.22331/q-2025-10-13-1884}
}

@article{Kobayashi2026ErasureNeutralAtom,
   title={Erasure-Tolerance Scheme for the Surface Codes on Neutral Atom Quantum Computers},
   volume={7},
   ISSN={2689-1808},
   journal={IEEE Transactions on Quantum Engineering},
   publisher={Institute of Electrical and Electronics Engineers (IEEE)},
   author={Kobayashi, Fumiyoshi and Nagayama, Shota},
   year={2026},
   pages={1–13} }

@article{gidney2021stim,
  doi = {10.22331/q-2021-07-06-497},
  url = {https://doi.org/10.22331/q-2021-07-06-497},
  title = {Stim: a fast stabilizer circuit simulator},
  author = {Gidney, Craig},
  journal = {{Quantum}},
  issn = {2521-327X},
  publisher = {{Verein zur F{\"{o}}rderung des Open Access Publizierens
                in den Quantenwissenschaften}},
  volume = 5,
  pages = 497,
  month = jul,
  year = 2021
}

@misc{pymatchingv2,
  author = {Higgott, Oscar and Gidney, Craig},
  title = {PyMatching v2},
  year = {2022},
  publisher = {GitHub},
  journal = {GitHub repository},
  howpublished = {\url{https://github.com/oscarhiggott/PyMatching}}
}

@article{miles2026qubitsyndromemeasurementshigh,
  title={Qubit syndrome measurements with a high fidelity Rb-Cs Rydberg gate},
  author={Miles, J and Lichtman, MT and Scott, AM and Scott, J and Norrell, SA and Bedalov, MJ and Belknap, DA and Cole, DC and Eubanks, SY and Gillette, M and others},
  journal={arXiv:2603.13492},
  year={2026},
  url={https://arxiv.org/abs/2603.13492}, 
}

@inproceedings{Viszlai2025interleaved,
  title={Interleaved Logical Qubits in Atom Arrays},
  author={Viszlai, Joshua and Lin, Sophia and Dangwal, Siddharth and Bradley, Conor and Ramesh, Vikram and Baker, Jonathan and Bernien, Hannes and Chong, Frederic T},
  booktitle={2025 IEEE International Symposium on High Performance Computer Architecture (HPCA)},
  pages={261--274},
  year={2025},
  organization={IEEE}
}

@article{Ramamoorthy1977ClasssicalPipelining,
author = {Ramamoorthy, C. V. and Li, H. F.},
title = {Pipeline Architecture},
year = {1977},
issue_date = {March 1977},
publisher = {Association for Computing Machinery},
address = {New York, NY, USA},
volume = {9},
number = {1},
issn = {0360-0300},
url = {https://doi.org/10.1145/356683.356687},
doi = {10.1145/356683.356687},
journal = {ACM Comput. Surv.},
month = mar,
pages = {61–102},
numpages = {42}
}

@article{Patomaki2024PipelinedQPUSiliconQubit,
	author = {Patom{\"a}ki, S. M. and Gonzalez-Zalba, M. F. and Fogarty, M. A. and Cai, Z. and Benjamin, S. C. and Morton, J. J. L.},
	journal = {npj Quantum Inf.},
	number = {1},
	pages = {31},
	title = {Pipeline quantum processor architecture for silicon spin qubits},
	volume = {10},
	year = {2024},
    url = {https://doi.org/10.1038/s41534-024-00823-y}
}

@inproceedings{Wang2024FPQA,
  title={Q-pilot: Field programmable qubit array compilation with flying ancillas},
  author={Wang, Hanrui and Tan, Daniel Bochen and Liu, Pengyu and Liu, Yilian and Gu, Jiaqi and Cong, Jason and Han, Song},
  booktitle={Proceedings of the 61st ACM/IEEE Design Automation Conference},
  pages={1--6},
  year={2024}
}

@inproceedings{LSQCA,
  title={Lsqca: Resource-efficient load/store architecture for limited-scale fault-tolerant quantum computing},
  author={Kobori, Takumi and Suzuki, Yasunari and Ueno, Yosuke and Tanimoto, Teruo and Todo, Synge and Tokunaga, Yuuki},
  booktitle={2025 IEEE International Symposium on High Performance Computer Architecture (HPCA)},
  pages={304--320},
  year={2025},
  organization={IEEE}
}

@article{FLASQ,
  title={The FLuid Allocation of Surface code Qubits (FLASQ) cost model for early fault-tolerant quantum algorithms},
  author={Huggins, William J and Khattar, Tanuj and Xu, Amanda and Harrigan, Matthew and Kang, Christopher and Low, Guang Hao and Fowler, Austin and Rubin, Nicholas C and Babbush, Ryan},
  journal={arXiv:2511.08508},
  year={2025},
  url = {https://doi.org/10.48550/arXiv.2511.08508}
}

@article{thaker2006quantummemoryhierarchiesefficient,
        author = {Thaker, Darshan D. and Metodi, Tzvetan S. and Cross, Andrew W. and Chuang, Isaac L. and Chong, Frederic T.},
        title = {Quantum Memory Hierarchies: Efficient Designs to Match Available Parallelism in Quantum Computing},
        year = {2006},
        publisher = {Association for Computing Machinery},
        volume = {34},
        number = {2},
        issn = {0163-5964},
        journal = {SIGARCH Comput. Archit. News},
        pages = {378–390},
        numpages = {13},
      url={https://doi.org/10.1145/1150019.1136518}, 
}

@article{Giovannetti2008QRAM,
  title = {Quantum Random Access Memory},
  author = {Giovannetti, Vittorio and Lloyd, Seth and Maccone, Lorenzo},
  journal = {Phys. Rev. Lett.},
  volume = {100},
  issue = {16},
  pages = {160501},
  numpages = {4},
  year = {2008},
  month = {Apr},
  publisher = {American Physical Society},
  doi = {10.1103/PhysRevLett.100.160501},
  url = {https://link.aps.org/doi/10.1103/PhysRevLett.100.160501}
}

@article{cesa2025fasterrorcorrectablequantumram,
      title={Fast and Error-Correctable Quantum RAM},
      author={Cesa, Francesco and Bernien, Hannes and Pichler, Hannes},
      journal={arXiv:2503.19172},
      year={2025},
      url={https://arxiv.org/abs/2503.19172}
}

@inproceedings{Stein2023HeterogeneousArchitecture,
  title={Hetarch: Heterogeneous microarchitectures for superconducting quantum systems},
  FULLauthor = {Stein, Samuel and Sussman, Sara and Tomesh, Teague and Guinn, Charles and Tureci, Esin and Lin, Sophia Fuhui and Tang, Wei and Ang, James and Chakram, Srivatsan and Li, Ang and Martonosi, Margaret and Chong, Fred and Houck, Andrew A. and Chuang, Isaac L. and Demarco, Michael},
  author={Stein, Samuel and Sussman, Sara and Tomesh, Teague and Guinn, Charles and Tureci, Esin and Lin, Sophia Fuhui and Tang, Wei and Ang, James and Chakram, Srivatsan and Li, Ang and others},
  booktitle={Proceedings of the 56th Annual IEEE/ACM International Symposium on Microarchitecture},
  pages={539--554},
  year={2023}
}

@article{yang2026spacetimeefficienthardwarecompatiblecomplexquantum,
  title={Spacetime-Efficient and Hardware-Compatible Complex Quantum Logic Units in qLDPC Codes},
  author={Yang, Willers and Chadwick, Jason and Teo, Mariesa H and Viszlai, Joshua and Chong, Fred},
  journal={arXiv:2602.14273},
  year={2026},
  url={https://arxiv.org/abs/2602.14273}, 
}

@article{Kitaev2005MagicStateDistillation,
  title = {Universal quantum computation with ideal Clifford gates and noisy ancillas},
  author = {Bravyi, Sergey and Kitaev, Alexei},
  journal = {Phys. Rev. A},
  volume = {71},
  issue = {2},
  pages = {022316},
  numpages = {14},
  year = {2005},
  month = {Feb},
  publisher = {American Physical Society},
  doi = {10.1103/PhysRevA.71.022316},
  url = {https://link.aps.org/doi/10.1103/PhysRevA.71.022316}
}

@book{steck2020QuantumOptics,
    title={Quantum and atom optics},
    author={Steck, Daniel A},
    year={revision 0.16.7, 2026},
    publisher = {available online at \href{http://steck.us/teaching}{http://steck.us/teaching}},
}

@article{Einwohner1976GraphBasedRotatingWaveApproximation,
  title = {Analytical solutions for laser excitation of multilevel systems in the rotating-wave approximation},
  author = {Einwohner, T. H. and Wong, J. and Garrison, J. C.},
  journal = {Phys. Rev. A},
  volume = {14},
  issue = {4},
  pages = {1452--1456},
  numpages = {0},
  year = {1976},
  month = {Oct},
  publisher = {American Physical Society},
  doi = {10.1103/PhysRevA.14.1452},
  url = {https://link.aps.org/doi/10.1103/PhysRevA.14.1452}
}

@article{LAMBERT20261Qutip5,
	author = {Lambert, Neill and Gigu{\`e}re, Eric and Menczel, Paul and Li, Boxi and Hopf, Patrick and Su{\'a}rez, Gerardo and Gali, Marc and Lishman, Jake and Gadhvi, Rushiraj and Agarwal, Rochisha and others},
	journal = {Phys. Rep.},
	pages = {1-62},
	title = {QuTiP 5: The Quantum Toolbox in Python},
	volume = {1153},
	year = {2026},
    url = {https://doi.org/10.1016/j.physrep.2025.10.001}
}

@book{Bransden2003AtomBook,
  title={Physics of atoms and molecules},
  author={Bransden, Brian Harold and Joachain, Charles Jean},
  year={2003},
  publisher={Pearson Education India}
}

@phdthesis{Martin2013SrClock,
  author  = {Michael J. Martin},
  title   = {Quantum Metrology and Many-Body Physics: Pushing the
Frontier of the Optical Lattice Clock},
  school  = {University of Colorado, Department of Physics},
  year = {2013},
  type = {Ph.D. thesis}
}

@article{Hohn2026YbTuneOut,
  title = {Determining the ${^3}P_0$ Excited-State Tune-Out Wavelength of $^{174}\mathrm{Yb}$ in a Triple-Magic Lattice},
  author = {H\"ohn, Tim O. and Villela, Ren\'e A. and Zu, Er and Bezzo, Leonardo and Kroeze, Ronen M. and Aidelsburger, Monika},
  journal = {PRX Quantum},
  volume = {7},
  issue = {1},
  pages = {010303},
  numpages = {8},
  year = {2026},
  month = {Jan},
  publisher = {American Physical Society},
  doi = {10.1103/32q9-j82c},
  url = {https://link.aps.org/doi/10.1103/32q9-j82c}
}

@article{kroeze2026171ybreferencedata,
  title={$^{171}\mathrm{Yb}$ Reference Data},
  author={Kroeze, Ronen M and Kristensen, Sofus Laguna and Pucher, Sebastian},
  journal={arXiv:2509.04416},
  year={2025},
  url={https://arxiv.org/abs/2509.04416}, 
}

@article{moore2023photon,
  title={Photon scattering errors during stimulated Raman transitions in trapped-ion qubits},
  author={Moore, ID and Campbell, WC and Hudson, ER and Boguslawski, MJ and Wineland, DJ and Allcock, DTC},
  journal={Phys. Rev. A},
  volume={107},
  number={3},
  pages={032413},
  year={2023},
  publisher={APS},
  url = {https://doi.org/10.1103/PhysRevA.107.032413}
}

@article{RamanError,
  title={Errors in trapped-ion quantum gates due to spontaneous photon scattering},
  author = {Ozeri, R. and Itano, W. M. and Blakestad, R. B. and Britton, J. and Chiaverini, J. and Jost, J. D. and Langer, C. and Leibfried, D. and Reichle, R. and Seidelin, S. and Wesenberg, J. H. and Wineland, D. J.},
  journal={Phys. Rev. A},
  volume={75},
  number={4},
  pages={042329},
  year={2007},
  publisher={APS},
  url = {https://doi.org/10.1103/PhysRevA.75.042329}
}

@article{RayleighError,
  title={Decoherence due to elastic Rayleigh scattering},
  author={Uys, Hermann and Biercuk, Michael J and VanDevender, Aaron P and Ospelkaus, Christian and Meiser, Dominic and Ozeri, Roee and Bollinger, John J},
  journal={Phys. Rev. Lett.},
  volume={105},
  number={20},
  pages={200401},
  year={2010},
  publisher={APS},
  url = {https://doi.org/10.1103/PhysRevLett.105.200401}
}

@article{Magesan2012RBTheory,
  title = {Characterizing quantum gates via randomized benchmarking},
  author = {Magesan, Easwar and Gambetta, Jay M. and Emerson, Joseph},
  journal = {Phys. Rev. A},
  volume = {85},
  issue = {4},
  pages = {042311},
  numpages = {16},
  year = {2012},
  month = {Apr},
  publisher = {American Physical Society},
  doi = {10.1103/PhysRevA.85.042311},
  url = {https://link.aps.org/doi/10.1103/PhysRevA.85.042311}
}

@article{Dorscher2018ScatteringRate,
  title = {Lattice-induced photon scattering in an optical lattice clock},
  author = {D\"orscher, S\"oren and Schwarz, Roman and Al-Masoudi, Ali and Falke, Stephan and Sterr, Uwe and Lisdat, Christian},
  journal = {Phys. Rev. A},
  volume = {97},
  issue = {6},
  pages = {063419},
  numpages = {9},
  year = {2018},
  month = {Jun},
  publisher = {American Physical Society},
  doi = {10.1103/PhysRevA.97.063419},
  url = {https://link.aps.org/doi/10.1103/PhysRevA.97.063419}
}

@article{Nowak2008QuantumTrajectory,
	author = {Ryan Nowak and James P. Clemens},
	journal = {J. Opt. Soc. Am. B},
	month = {Apr},
	number = {4},
	pages = {564--570},
	title = {Quantum trajectory theory of superradiant emission from randomly distributed atomic samples},
	volume = {25},
	year = {2008},
    url = {https://doi.org/10.1364/JOSAB.25.000564}
}

@article{Clemens2003CollectiveSpontaneousEmission,
  title = {Collective spontaneous emission from a line of atoms},
  author = {Clemens, J. P. and Horvath, L. and Sanders, B. C. and Carmichael, H. J.},
  journal = {Phys. Rev. A},
  volume = {68},
  issue = {2},
  pages = {023809},
  numpages = {19},
  year = {2003},
  month = {Aug},
  publisher = {American Physical Society},
  doi = {10.1103/PhysRevA.68.023809},
  url = {https://link.aps.org/doi/10.1103/PhysRevA.68.023809}
}

@article{Daley2014QuantumTrajectory,
   title={Quantum trajectories and open many-body quantum systems},
   volume={63},
   ISSN={1460-6976},
   url={http://dx.doi.org/10.1080/00018732.2014.933502},
   DOI={10.1080/00018732.2014.933502},
   number={2},
   journal={Adv. Phys.},
   publisher={Informa UK Limited},
   author={Daley, Andrew J.},
   year={2014},
   month=mar, pages={77–149} 
}

@article{Ghosh2012TwirlingDecoherence,
  title = {Surface code with decoherence: An analysis of three superconducting architectures},
  author = {Ghosh, Joydip and Fowler, Austin G. and Geller, Michael R.},
  journal = {Phys. Rev. A},
  volume = {86},
  issue = {6},
  pages = {062318},
  numpages = {12},
  year = {2012},
  month = {Dec},
  publisher = {American Physical Society},
  doi = {10.1103/PhysRevA.86.062318},
  url = {https://link.aps.org/doi/10.1103/PhysRevA.86.062318}
}

@article{Pucher2024FineStructureSr,
  title = {Fine-Structure Qubit Encoded in Metastable Strontium Trapped in an Optical Lattice},
  author = {Pucher, S. and Kl\"usener, V. and Spriestersbach, F. and Geiger, J. and Schindewolf, A. and Bloch, I. and Blatt, S.},
  journal = {Phys. Rev. Lett.},
  volume = {132},
  issue = {15},
  pages = {150605},
  numpages = {6},
  year = {2024},
  month = {Apr},
  publisher = {American Physical Society},
  doi = {10.1103/PhysRevLett.132.150605},
  url = {https://link.aps.org/doi/10.1103/PhysRevLett.132.150605}
}

@article{unnikrishnan2024coherent,
  title={Coherent control of the fine-structure qubit in a single alkaline-earth atom},
  author={Unnikrishnan, Govind and Ilzh{\"o}fer, Philipp and Scholz, Achim and H{\"o}lzl, Christian and G{\"o}tzelmann, Aaron and Gupta, Ratnesh Kumar and Zhao, Jiachen and Krauter, Jennifer and Weber, Sebastian and Makki, Nastasia and others},
  journal={Phys. Rev. Lett.},
  volume={132},
  number={15},
  pages={150606},
  year={2024},
  publisher={APS},
  url = {https://doi.org/10.1103/PhysRevLett.132.150606}
}

@article{Madjarov2020OpticalQubit,
	author = {Madjarov, Ivaylo S. and Covey, Jacob P. and Shaw, Adam L. and Choi, Joonhee and Kale, Anant and Cooper, Alexandre and Pichler, Hannes and Schkolnik, Vladimir and Williams, Jason R. and Endres, Manuel},
	journal = {Nat. Phys.},
	number = {8},
	pages = {857--861},
	title = {High-fidelity entanglement and detection of alkaline-earth Rydberg atoms},
	volume = {16},
	year = {2020},
    url = {https://www.nature.com/articles/s41567-020-0903-z}
}

@article{Young2020clockqubit,
	author = {Young, Aaron W. and Eckner, William J. and Milner, William R. and Kedar, Dhruv and Norcia, Matthew A. and Oelker, Eric and Schine, Nathan and Ye, Jun and Kaufman, Adam M.},
	journal = {Nature},
	number = {7838},
	pages = {408--413},
	title = {Half-minute-scale atomic coherence and high relative stability in a tweezer clock},
	volume = {588},
	year = {2020},
    url ={https://www.nature.com/articles/s41586-020-3009-y}
}

@article{Allcock2021omgIon,
    author = {Allcock, D. T. C. and Campbell, W. C. and Chiaverini, J. and Chuang, I. L. and Hudson, E. R. and Moore, I. D. and Ransford, A. and Roman, C. and Sage, J. M. and Wineland, D. J.},
    title = {omg blueprint for trapped ion quantum computing with metastable states},
    journal = {Appl. Phys. Lett.},
    volume = {119},
    number = {21},
    pages = {214002},
    year = {2021},
    month = {11},
    issn = {0003-6951},
    doi = {10.1063/5.0069544},
    url = {https://doi.org/10.1063/5.0069544},
}
